\documentclass[aps,prb,floatfix,amsmath,amssymb,eqsecnum,nofootinbib,twocolumn,superscriptaddress]{revtex4}

\usepackage{amsmath,amssymb,dsfont,graphicx,psfrag}
\usepackage{epsfig}
\usepackage{graphicx}
\usepackage{epstopdf}
\usepackage{tikz}
\usetikzlibrary{decorations.markings}
\tikzset{test/.style n args={3}{
		postaction={
			decorate,
			decoration={
				markings,
				mark=between positions 0 and \pgfdecoratedpathlength step 0.5pt with {
					\pgfmathsetmacro\myval{multiply(
						divide(
						\pgfkeysvalueof{/pgf/decoration/mark info/distance from start}, \pgfdecoratedpathlength
						),
						100
						)};
					\pgfsetfillcolor{#3!\myval!#2};
					\pgfpathcircle{\pgfpointorigin}{#1};
					\pgfusepath{fill};}
}}}}
\usepackage{xcolor}
\usetikzlibrary{patterns}
\usetikzlibrary{calc}		
	
\def\beq{\begin{equation}}
\def\eeq{\end{equation}}

\begin{document}
\title{Quantum Thermal Hall effect of chiral spinons on a Kagome strip}

\author{Pavel Tikhonov}
\affiliation{Department of Physics, Bar-Ilan University, Ramat-Gan 52900, Israel}

\author{Efrat Shimshoni}
\affiliation{Department of Physics, Bar-Ilan University, Ramat-Gan 52900, Israel}

\begin{abstract}
We develop a theory for the thermal Hall coefficient in a spin-$\frac{1}{2}$ system on a strip of Kagome lattice, where a chiral spin-interaction term is present. To this end, we model the Kagome strip as a three-leg $XXZ$ spin-ladder, and use Bosonization to derive a low-energy theory for the spinons in this system. Introducing further a Dzyaloshinskii-Moriya interaction ($D$) and a tunable magnetic field ($B$), we identify three distinct $B$-dependent quantum phases: a valence-bond crystal (VBC), a ``metallic" spin liquid (MSL) and a chiral spin liquid (CSL). In the presence of a temperature difference $\Delta T$ between the top and the bottom edges of the strip, we evaluate the net heat current $J_h$ along the strip, and consequently the thermal Hall conductivity $\kappa_{xy}$. We find that the VBC-MSL-CSL transitions are accompanied by a pronounced qualitative change in the behavior of $\kappa_{xy}$ as a function of $B$. In particular, analogously to the quantum Hall effect, $\kappa_{xy}$ in the CSL phase exhibits a quantized plateau centered around a commensurate value of the spinon filling factor $\nu_s\propto B/D$.
\end{abstract}
\maketitle

\section{Introduction}
Magnetic compounds dominated by spin-$\frac{1}{2}$ degrees of freedom which are subject to competing interactions provide a fascinating platform for the potential realization of exotic quantum phases. A prominent example is the case of an anti-ferromagnetic (AFM) Heisenberg magnet on a geometrically frustrated lattice such as the Kagome or Pyrochlore structures, where a magnetically ordered ground state with well-defined local spin orientation is not favorable. A possible consequence is the formation of a state of matter dubbed a quantum spin liquid (SL), a term first introduced by Anderson \cite{Anderson1973} along with a concrete example: the resonating valence bond (RVB) state. An appealing property of such state is that it exhibits an extreme case of spin-charge separation in strongly correlated electron systems, where an electric insulator (in which the charge sector is completely frozen) supports ``electron-like" low-energy fractionalized excitations (spinons). Hence in the last decades, the search for SL phases in various quantum spin systems has motivated considerable theoretical and experimental work \cite{balents-2010,SavaryBalents2017}.

Convincing evidence for the existence of SL phases in realistic materials is, however, rather scarce. A primary challenge is that such a state is extremely sensitive to the fine balance between competing spin-exchange interactions \cite{Starykh}. These can favor alternative ground states which break translational symmetry and possess a local order parameter, such as spin density wave or a valence bond crystal\cite{Iwase1996, Azuma1994, Kageyama1999} (VBC) -- an ordered pattern where singlets are formed on particular bonds in the lattice. In certain models, a SL state was found to be confined to a fine-tuned critical point \cite{NT,Senthil2004}. Specifically for Kagome AFM, numerical studies are highly challenged by finite system size; thus far, despite applications of powerful methods, they have not lead to a clear consensus on the nature of ground state \cite{DMRG,Assa2013}.

Experimentally, conclusively identifying a SL state is also a challenge. Because of its liquid nature, thermodynamic measurements such as magnetic susceptibility only show the {\em absence} of magnetic order down to low temperatures \cite{Hiroi2001,Ofer2006,Helton2007,Olariu,Yamashita2008,Okamoto2009,Matan}. As an alternative probe, heat transport measurements give access to neutral low-energy excitations, and provided some evidence for the presence of spinons in SL-candidate materials \cite{Yamashita2009}. Detecting magneto-thermal transport under application of a magnetic field can serve an effective mean to disentangle their contribution from the phonon background. Interestingly, in certain magnetic insulating compounds (involving heavy elements) such measurements indicated a finite transverse component, i.e. a thermal Hall conductivity\cite{HirschbergerChinellLeeOng2015,HirschbergerKrizanCavaOng2015} $\kappa_{xy}$. This suggests the presence of chiral spin-interaction terms, generated due to the enhanced spin-orbit coupling.

A pronounced role of chiral interactions provides the basis for a unique species of SL -- a Chiral SL (CSL) -- which does possess a local order parameter:
the expectation value of a "3-spin" operator ${\bf S}_{i}\cdot\left({\bf S}_{j}\times{\bf S}_{k}\right)$, where $i$, $j$, $k$ belong to a triangle of a given lattice\cite{WenWilczekZee1989, Baskaran1989}. Most prominently, the CSL provides an analogue of the fractional quantum Hall effect\cite{KalmeyerLaughlin1987, KalmeyerLaughlin1989} (FQHE) in a charge-insulating electronic system, where spinons are subjected to a fictitious flux on triangular plaquettes. More recent theoretical studies \cite{EranSela2015,Moessner2015,TobiasMeng2015} have confirmed the emergence of such phase in specific lattice models. The anticipated hallmark of such a state is the quantization of thermal Hall conductivity in units of $\frac{\pi}{6}\frac{k_{_B}^2}{\hbar}T$ (with $T$ the temperature).

Lately, progress in the experimental search for CSL behavior has been achieved by studies of Ir/Ru compounds, which serve as potential realizations of the Kitaev model \cite{Kitaev2006,KitaevSL}. Remarkably, this model possesses an exact solution in two-dimensions (2D) by mapping to free Fermions, and predicts fractionalized quasi-particles of which the gapless type are Majorana Fermions. Their expected signature is a fractional thermal Hall effect: $\kappa_{xy}=q\frac{\pi}{6}\frac{k_{_B}^2}{\hbar}T$ with $q=1/2$. A pioneering recent measurement \cite{Kasahara2018} has confirmed the existence of a quantized plateau at this value in $\alpha$-RuCl$_3$, though confined to a narrow range of the applied magnetic field $B$. Additional evidence for the presence of a chiral order in these materials is provided by magnetic torque measurements \cite{torque1,torque2}.

While the above mentioned experimental results, as well as the earlier thermal Hall measurements \cite{HirschbergerChinellLeeOng2015,HirschbergerKrizanCavaOng2015}, provide encouraging evidence for chiral spin excitations, certain crucial features of the data call for further theoretical investigation. In particular, the rather complex and non-monotonic $B$-dependence exhibited by $\kappa_{xy}$ can not be fully explained by means of an ideal spin model \cite{HyunyongJungPatrick,VinklerAvivRosch2018}. Moreover, it reflects the sensitivity of a CSL phase (if such exists) to system parameters. Motivated by these observations, our present work addresses a tractable minimal model which allows to systematically explore the possible quantum phases of chiral spin systems, their evolution with variations in a tunable parameter such as the external field $B$, and their manifestation in the thermal Hall effect.

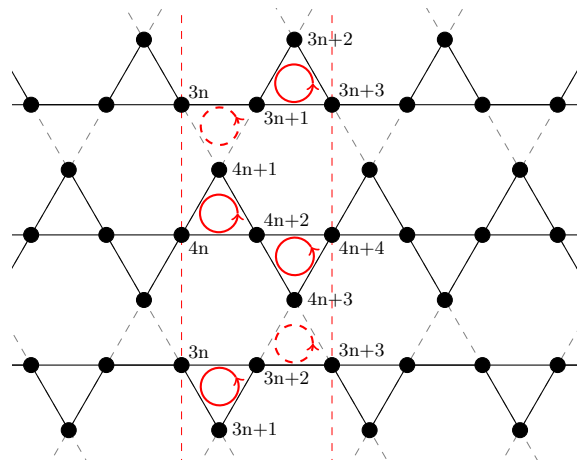
\begin{figure}
\begin{tikzpicture}[scale=1]
	\clip (-1.25,-1.25) rectangle (6.25,4.75);
	\coordinate (a1) at (2,0);
	\coordinate (a2) at (1,{sqrt(3)});	
	\coordinate (d1) at (0,0);
	\coordinate (d2) at (1,0);	
	\coordinate (d3) at (0.5,{0.5*sqrt(3)});
	\coordinate (c) at ($0.33*(d1)+0.33*(d2)+0.33*(d3)$);
		\draw [style=help lines,dashed,shorten >=-10cm,shorten <=-10cm,red] ($0.5*(a1)$) -- ($2*(a2)-0.5*(a1)$) {};
		\draw [style=help lines,dashed,shorten >=-10cm,shorten <=-10cm,red] ($0.5*(a1)+1*(a1)$) -- ($2*(a2)+0.5*(a1)$) {};
	
	\foreach \n in {-3,-2,...,3}{
        \draw [shorten >=-10cm,shorten <=-10cm] ($\n*(a2)$) -- ($(a1)+\n*(a2)$);
		\draw [style=help lines,dashed,shorten >=-10cm,shorten <=-10cm] ($\n*(a1)$) -- ($(a2)+\n*(a1)$) {};
		\draw [style=help lines,dashed,shorten >=-10cm,shorten <=-10cm] ($0.5*(a1)+0.5*\n*(a2)+0.5*\n*(a1)$) -- ($0.5*(a2)+0.5*\n*(a2)+0.5*\n*(a1)$) {};
    }
	\foreach \n in {-3,-2,...,4}{
		\foreach \m in {-3,-2,...,4}{
        	\node[draw,circle,inner sep=2pt,fill] at ($\n*(a1)+\m*(a2)+(d1)$) {};
        	\node[draw,circle,inner sep=2pt,fill] at ($\n*(a1)+\m*(a2)+(d2)$) {};
        	\node[draw,circle,inner sep=2pt,fill] at ($\n*(a1)+\m*(a2)+(d3)$) {};
	}
		}
	\node[scale=0.8, above right] at ($0.5*(a1)$) {3n};
	\node[scale=0.8, right=1mm ] at ($(a1)-0.5*(a2)$) {3n+1};
	\node[scale=0.8, below right] at ($(a1)$) {3n+2};
	\node[scale=0.8, above right] at ($0.5*(a1)+(a1)$) {3n+3};
	\draw ($(0,0)-(a1)$) -- ($(0,0)-0.5*(a1)$) -- ($(0,0)-0.5*(a2)$) -- (0,0) -- ($0.5*(a1)$) -- ($(a1)-0.5*(a2)$) -- ($(a1)$) -- ($0.5*(a1)+(a1)$) -- ($(a1)+(a1)-0.5*(a2)$) -- ($(a1)+(a1)$) -- ($0.5*(a1)+2*(a1)$) -- ($3*(a1)-0.5*(a2)$) -- ($3*(a1)$)-- ($3.5*(a1)$) ;
	\node[scale=0.8, below right] at ($(a2)$) {4n};
	\node[scale=0.8, right=1mm ] at ($(a2)+0.5*(a2)$) {4n+1};
	\node[scale=0.8, above right] at ($(a2)+0.5*(a1)$) {4n+2};
	\node[scale=0.8, right=1mm] at ($0.5*(a2)+(a1)$) {4n+3};
	\node[scale=0.8, below right] at ($(a2)+(a1)$) {4n+4};
	\draw ($0.5*(a2)-(a1)$) -- ($(a2)-(a1)$) -- ($(a2)+0.5*(a2)-(a1)$) -- ($(a2)-0.5*(a1)$) -- ($0.5*(a2)$) -- ($(a2)$) -- ($(a2)+0.5*(a2)$) -- ($(a2)+0.5*(a1)$) -- ($0.5*(a2)+(a1)$) -- ($(a2)+(a1)$) -- ($(a2)+0.5*(a2)+(a1)$) -- ($(a2)+1.5*(a1)$) -- ($0.5*(a2)+2*(a1)$) -- ($(a2)+2*(a1)$) -- ($(a2)+0.5*(a2)+2*(a1)$) -- ($(a2)+2.5*(a1)$) -- ($0.5*(a2)+3*(a1)$);
	\node[scale=0.8, above right] at ($2*(a2)-0.5*(a1)$) {3n};
	\node[scale=0.8, below right] at ($2*(a2)$) {3n+1};
	\node[scale=0.8, right=1mm] at ($2.5*(a2)$) {3n+2};
	\node[scale=0.8, above right] at ($2*(a2)+0.5*(a1)$) {3n+3};
	
	\draw ($2.5*(a2)-2*(a1)$) -- ($2*(a2)-1.5*(a1)$) -- ($2*(a2)-(a1)$) -- ($2.5*(a2)-(a1)$) -- ($2*(a2)-0.5*(a1)$) -- ($2*(a2)$) -- ($2.5*(a2)$) -- ($2*(a2)+0.5*(a1)$) -- ($2*(a2)+(a1)$) -- ($2.5*(a2)+(a1)$) -- ($2*(a2)+1.5*(a1)$) -- ($2*(a2)+2*(a1)$) -- ($2.5*(a2)+2*(a1)$) -- ($2*(a2)+2.5*(a1)$)  ;
	\foreach \n in {1}{        		
		\foreach \m in {0}{
			\draw[thick,dashed,red,-{>[scale=2.0]}] ([shift=(0:0.25cm)]$\n*(a1)+\m*(a2)+(c)$) arc (0:360:0.25cm);
			\draw[thick,red,-{>[scale=2.0]}] ([shift=(30:0.25cm)]$\n*(a1)+\m*(a2)-(c)$)arc (30:390:0.25cm);
     }
     	}
	\draw[thick,red,-{>[scale=2.0]}] ([shift=(0:0.25cm)]$0*(a1)+1*(a2)+(c)$) arc (0:360:0.25cm);
	\foreach \n in {0}{        		
		\foreach \m in {2}{
			\draw[thick,red,-{>[scale=2.0]}] ([shift=(0:0.25cm)]$\n*(a1)+\m*(a2)+(c)$) arc (0:360:0.25cm);
			\draw[thick,dashed,red,-{>[scale=2.0]}] ([shift=(30:0.25cm)]$\n*(a1)+\m*(a2)-(c)$)arc (30:390:0.25cm);
     }
     	}
	\draw[thick,red,-{>[scale=2.0]}] ([shift=(30:0.25cm)]$1*(a1)+1*(a2)-(c)$)arc (30:390:0.25cm);
\end{tikzpicture}
\caption{(Color online) Structure of the Kagome strip. Red dashed vertical lines represent the boundaries of unit cell $n$; black dots represent localized spins, black lines spin-spin exchange bonds and red circles the chiral interaction, where solid lines denote intra-chain interactions and dashed lines inter-chain coupling.}
\label{fig:ladders}
\end{figure}

To this end, in this paper we investigate a quasi one-dimensional (1D) model for a spin-$\frac{1}{2}$ system on a strip of Kagome lattice, in the presence of a 3-spin chiral interaction, a magnetic field $B$ and a Dzyaloshinskii-Moriya (DM) interaction $D$. To allow for further tunability, we incorporate anisotropy of the exchange interactions which break both $SU(2)$-symmetry and the lattice symmetry (see Fig. \ref{fig:ladders}). This enables a treatment of the model in terms of weakly coupled $XXZ$ spin-$\frac{1}{2}$ chains and subsequent application of Bosonization, which facilitates the analysis of the phase diagram. We then derive the thermal Hall coefficient $\kappa_{xy}$ by evaluating the net heat current along the strip in response to temperature gradient across the transverse direction, and analyze its dependence on $B$ and $T$ in each of the phases.

\begin{figure}
	
\begin{tikzpicture}

\draw[->] (0,0) -- (5,0) node[below] {$B$};
\draw[->] (0,0) -- (0,4) node[left] {$\frac{\kappa_{xy}}{T}$};

\draw[style=help lines,dashed] (0,3) -- (3.5,3);
\node[left] at (0,3) {$\frac{\pi}{6}$};
\draw[style=help lines,dashed] (3.625,3) -- (3.625,0);
\node[below] at (3.625,0) {$B_{D}$};
\draw[scale=1,domain=0.01:1.5,smooth,variable=\x, blue,  line width=0.7] plot ((\x,{\x*(0.21)*(1/((exp(-(\x-0.75)/(0.25))+1))});
\draw[violet, line width=0.7] (1.5,0.3) -- (2.75,0.55);
\draw[scale=1, test={0.4pt}{violet}{red}] (2.75, 0.55)..controls(3.125, 0.625) and (3.125, 3.0)..(3.5, 3.0);
\draw[red, line width=0.7] (3.5,3.0) -- (3.75,3.0);
\draw[scale=1, dashed, test={0.4pt}{red}{violet}] (3.75, 3)..controls(4.125, 3) and (4.125, 0.825)..(4.5, 0.9);
\draw[violet, line width=0.7] (4.5,0.9) -- (5,1);
\end{tikzpicture}
	\caption{(Color online) Schematic behavior of $\kappa_{xy}/T$ (in units of $k_{B}^2/\hbar$) as a function of magnetic field $B$; here $B_D$ is the value of $B$ obeying $k_B=\pm k_D/4$ (see text).}
\label{fig:IntroPlot}
\end{figure}
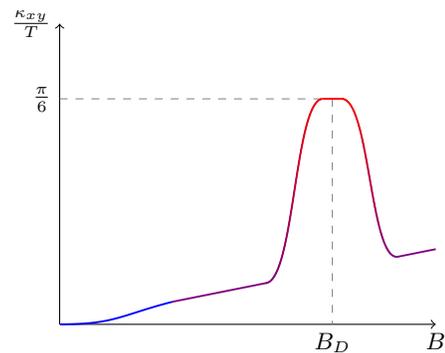

As a result of this analysis, we identify three distinct $B$-dependent quantum
phases. For low $B$ we obtain a VBC phase with a gap to spin excitations, which makes it a "spin-insulator" with activation behavior of the heat transport. By further increasing the field $B$ it is possible to reach a commensurability condition between the spinon density (dictated by $B$) and the "magnetic flux" (proportional to the DM coefficient $D$), which leads to the formation of CSL phase, in transparent analogy with FQHE states in electronic ladders \cite{FQHladders}. Its thermal Hall conductance $\kappa_{xy}$ exhibits a quantized plateau centered around the commensurate value of the field, $B_D\propto D$. Finally, the phase that is achieved for other values of $B$ is a plain SL we dub a metallic SL (MSL) formed by coupled Luttinger liquid channels, where the main contribution to $\kappa_{xy}$ results from inter-chain spinon scattering.  Our main result can be summarized by Fig. \ref{fig:IntroPlot}, where we schematically show $\kappa_{xy}$ as function of the magnetic field $B$ while driving a transitions from one phase to another (i.e., along a vertical cut through the phase diagram depicted in Fig. \ref{fig:RelevanceOfCosine}).

The paper is organized as follows: In Sec. \ref{sec:Model} we present the model of coupled spin chains on a Kagome strip, and identify the low-energy theory followed by analysis of its most dominant terms. In Sec. \ref{sec:PhaseDiagram} we analyze the phase diagram emanating from this effective theory. In Sec. \ref{sec:Thermal Hall Conductivity} we derive expressions for the heat current operator, and consequently for $\kappa_{xy}$ as function of $B$ and $T$ in each of the three phases. Finally, we present concluding remarks in Sec. \ref{sec:sum_remarks}. Throughout the paper, we use units where $\hbar=k_{B}=1$.

\section{The Model}
\label{sec:Model}
We consider a spin--$\frac{1}{2}$ system on a long strip of Kagome lattice with anisotropic exchange interactions as depicted in Fig. \ref{fig:ladders}, regarded as a $3$-leg ladder of meandering XXZ spin chains which are weakly coupled via the dashed bonds in the figure. We further assume that spin--orbit coupling in the underlying electronic system and an externally applied magnetic field $B$ (along the $\hat{z}$-axis of the spins) lead to explicit breaking of both time-reversal and parity symmetry. We hence include a Zeeman coupling to $B$, and introduce additional spin interactions including a Dzyaloshinskii-Moriya (DM) coupling to a vector $\vec{D}=D\hat{z}$, as well as a chiral $3$-spin ring-exchange interactions on all triangles. The Hamiltonian describing this system is
\begin{align}
\label{eq:H_original}
&H=\sum_{l=-1,0,1}H_{l}+H_{\perp} \; , \\
&H_{l}=\sum_{\left\langle i,j\right\rangle }J_{\parallel}^{z}S_{l,i}^{z}S_{l,j}^{z}+\frac{1}{2}J_{\parallel}^{xy}\left(S_{l,i}^{+}S_{l,j}^{-}+h.c.\right) \\
&+J_{\parallel}^{ch}\sum_{\left\{ i,j,k\right\} }\vec{S}_{l,i}\cdot\left(\vec{S}_{l,j}\times\vec{S}_{l,k}\right)\nonumber\\
&-B\sum_{i}S_{l,i}^{z}+D\sum_{\left[i,j\right]}\left(\vec{S}_{l,i}\times\vec{S}_{l,j}\right)_{z} \nonumber \\
\label{eq:H_perp_original}
&H_{\perp}=\sum_{\left\langle i,j\right\rangle }\sum_{\left\langle l,l'\right\rangle }J_{\perp}^{z}S_{l,i}^{z}S_{l',j}^{z}+\frac{1}{2}J_{\perp}^{xy}\left(S_{l,i}^{+}S_{l',j}^{-}+h.c.\right) \nonumber\\
&+\sum_{\left\langle l,l'\right\rangle }\Big\{J_{\perp}^{ch}\sum_{\left\{ i,j,k\right\} }\vec{S}_{l,i}\cdot\left(\vec{S}_{l',j}\times\vec{S}_{l',k}\right) \\
&+D\sum_{\left[i,j\right]}\left(\vec{S}_{l,i}\times\vec{S}_{l',j}\right)_{z}\Big\}\nonumber.
\end{align}
where $l$ denotes the chain index ($l=-1$, $l=0$ and $l=1$ denoting the bottom, middle and top chains, respectively); $\left<i,j\right>$ stands for nearest neighbor sites, $\left\lbrace i,j,k\right\rbrace$ for the corners of the same triangle, $\left[i,j\right]$ for sites connected by an edge of a triangle and $\left<l,l'\right>$ for adjacent chains. The actual choice of indexing by unit cell along the 1D periodic structure is given in Fig \ref{fig:ladders}. We assume the exchange coupling constant $J_{\parallel}^{\alpha}$, $J_{\perp}^{\alpha}$ to be anti-ferromagnetic, and that $J_{\parallel/\perp}^{ch}$ is proportional to the magnetic field. Also, for our construction to be valid we assume that the inter-chain couplings $J_{\perp}^{ch}$, $J_{\perp}^{xy}$ and $J_{\perp}^{z}$, as well as $J_{\parallel}^{ch}$, $B$ and $D$ are small relatively to $J_{\parallel}^{xy}$.

To properly bosonize the above Hamiltonian, we first perform a Jordan-Wigner (JW) transformation for each of the chains followed by bosonization of fermions, leading to the well known mapping of the spin operators in the continuum limit\cite{Giamarchi}
\begin{align}
\begin{split}
\label{eq:BosonizationMapping}
&S^{\pm}\left(x\right)\simeq\frac{1}{\sqrt{2\pi a}}e^{\mp i\tilde{\theta}_{l}}\left[\left(-\right)^{x}+\cos2\tilde{\phi_{l}}\right]
\\
&S_{l}^{z}\left(x\right)\simeq-\frac{1}{\pi}\partial_{x}\tilde{\phi_{l}}+\frac{1}{\pi a}\left(-\right)^{x}\cos2\tilde{\phi_{l}}.
\end{split}
\end{align}
Here $a$ is the lattice constant, $x=na$ with $n$ integer and the bosonic fields $\tilde{\phi_l}$, $\tilde{\theta_l}$ obey the canonical commutation relation
\begin{align}
\left[\tilde{\phi_{l}}\left(x'\right),\ \partial_{x}\tilde{\theta_{l}}\left(x\right)\right]=i\pi\delta\left(x-x'\right). \nonumber
\end{align}
Employing the mapping Eq. (\ref{eq:BosonizationMapping}) to obtain the continuum limit of $H$ [Eq. (\ref{eq:H_original})] should be carried out with some caution. Although the bosonization of spin-$\frac{1}{2}$ models can be found in the literature\cite{Giamarchi}, we present some details of the derivation for the less common terms in App. \ref{sec:Bosonizationon}.

The resulting low-energy representation of the Hamiltonian includes a quadratic part, where the $XXZ$ couplings on chain $l$ yield a Luttinger liquid (LL) with velocity $u_{l}$ and Luttinger parameter $K_{l}$. The magnetic field and DM coupling generate terms linear in the fields $\tilde{\theta_l}$ and $\tilde{\phi_l}$
\begin{align}
B\sum_{l=\pm1,0}\int\mathrm{d}x\frac{1}{\pi}\partial_{x}\tilde{\phi}_{l} +\frac{2D}{\pi}\int\mathrm{d}x\left(\partial_{x}\tilde{\theta}_{1}-\partial_{x}\tilde{\theta}_{-1}\right)
\end{align}
which can be absorbed into the definition of a new set of fields:
\begin{align}
\begin{split}
\label{eq:shift_trans}
&\theta_l=\tilde{\theta}_{l} + k^{D}_{l}x \\
&\phi_{l}=\tilde{\phi}_{l} + k^{B}_{l}x
\end{split}
\end{align}
where, for $J_{\parallel}^{ch},J_{\perp}^{ch},J_{\perp}^{\alpha}\ll J_{\parallel}^{xy}$, $k_{l}^{D}\simeq\frac{2D}{K_{l}u_{l}}l$ and $k_{l}^{B}\simeq\frac{K_{l}B}{u_{l}}$. We thus obtain for each chain $l$
\begin{align}
\label{eq:H_LowEnergy}
&H_{l}=H_{l}^{0}+H_{l}^{int} \\
&H_{l}^{0}=H_{l}^{LL}-g_{l}^{ch}\int\mathrm{d}x\partial_{x}\theta_{l}\partial_{x}\phi_{l}, \quad g_{l}^{ch}\equiv lg \nonumber \\
\label{eq:H_LL}
&H_{l}^{LL}=\frac{u_{l}}{2\pi}\int\mathrm{d}x\left\{ K_{l}\left(\partial_{x}\theta_{l}\right)^{2}+\frac{1}{K_{l}}\left(\partial_{x}\phi_{l}\right)^{2}\right\} \\
&H_{l}^{int}=\frac{2g_{l}}{\left(2\pi a\right)^{2}}\int \mathrm{d}x\cos\left(4\phi_l-4k^B_l x \right) \label{eq:H_cos},
\end{align}
in which $g=J_{\parallel}^{ch} \frac{8}{\pi^2}a$, $g_{\pm1}=2(J_{\parallel}^{xy}-J_{\parallel}^{z})a$, $g_{0}=2(2J_{\parallel}^{xy}-J_{\parallel}^{z})a$. In the perturbative regime ($|J_{\parallel}^{z}| \ll J_{\parallel}^{xy}$) the Luttinger liquid parameters have the following values
\begin{align}
\label{eq:LL_parameters}
\begin{split}
&K_{\pm1}^{2}\simeq\frac{1}{1+\frac{8}{3\pi}+\frac{4}{\pi}\Delta}\equiv K^{2}, \\
&K_{0}^{2}\simeq\frac{1}{1+\frac{4}{\pi}+\frac{4}{\pi}\Delta},\quad \Delta=\frac{J_{\parallel}^{z}}{J_{\parallel}^{xy}},
\end{split}\\
&u_l\simeq J^{xy}_{\parallel}a\frac{1}{K_l}, \quad u_{\pm1}\equiv u.
\end{align}
Note that the Luttinger parameters $K_l$ are reduced compared to the linear $XXZ$ chain because of next-nearest-neighbors coupling (e.g., the $3n+1$ and $3n+3$ sites in Fig. \ref{fig:ladders}). The coupling constants $g_l$ are positive for  $J^{z}_{\parallel}<J^{xy}_{\parallel}$, so that at $B=0$ (such that the oscillatory phase factor $k^D_l x$ vanishes) $H_{l}^{int}$ [Eq.(\ref{eq:H_cos})] is minimized by $2\phi_l=\pm\frac{\pi}{2}$. Therefore, when this term is relevant, it generates a dimerized state with $\langle S_{z}\rangle=0$.

The inter-chain coupling becomes
\begin{align}
\label{eq:H_perp}
&H_{\perp}=H_{\perp}^{0}+H^{ch} \\
\label{eq:H^0_perp}
&H_{\perp}^{0}=g_{\perp}^{z}\int\mathrm{d}x\left(\partial_{x}\phi_{1}+\partial_{x}\phi_{-1}\right)\partial_{x}\phi_{0}\\
&+g_{\perp,0}^{ch}\int\mathrm{d}x\left(\partial_{x}\theta_{1}-\partial_{x}\theta_{-1}\right)\partial_{x}\phi_{0} \nonumber \\
\label{eq:H^ch_perp}
&H_{\perp}^{ch}=\sum_{l=\pm1}\int\mathrm{d}x\frac{g_{\perp}^{ch}}{\left(2\pi a\right)^{2}}\Big\{\\
&+\cos\left(2\phi_{l}+2\phi_{0}-\theta_{l}+\theta_{0}-\Delta k_{-,l}x-\delta_{-,l}\right) \nonumber \\
&+\cos\left(2\phi_{l}+2\phi_{0}+\theta_{l}-\theta_{0}-\Delta k_{+,l}x-\delta_{+,l}\right)\Big\}, \nonumber
\end{align}
where
\begin{align}
\label{eq:CommensurateCond}
&\Delta k_{\pm,l}=4k_{B}\pm lk_{D}, \\ \nonumber
\begin{split}
&k_{B}\equiv\frac{1}{2}\left(k_{\pm1}^{B}+k_{0}^{B}\right)=\frac{1}{2}\left(\frac{K}{u}+\frac{K_{0}}{u_{0}}\right)B \\
&k_{D}\equiv lk_{l}^{D}=\frac{2D}{Ku}\; ,
\end{split}
\end{align}
$g^{z}_{\perp}=J^{z}_{\perp}\frac{2}{\pi^{2}}a$ and $g^{ch}_{\perp,0}=J^{ch}_{\perp}\frac{2}{\pi^{2}}a$; as $H_{\perp}^{ch}$ [Eq. (\ref{eq:H^ch_perp})] combines contributions from the last three terms of Eq. (\ref{eq:H_perp_original}) (see App. A), $g_{\perp}^{ch}$ and the constant phase shifts $\delta_{\pm,l}$ are functions of $J_{\perp}^{xy}$, $J_{\perp}^{ch}$, $D$ and $B$. In particular, such a term exists even if in the microscopic Hamiltonian, the bare chiral parameter $J_{\perp}^{ch}=0$. Additional contributions to the low-energy Hamiltonian, which are not capable of generating a mass, are ignored at this stage and will be discussed later in the paper.

Note that the chiral term Eq. (\ref{eq:H^ch_perp}) includes four terms which are typically frustrated due to the oscillation with wave-vector $\Delta k_{\pm}=4k_{B}\pm k_{D}$. Hence, this term may turn relevant only provided $k_B\simeq \pm k_D/4$. When either of these conditions on the field $B$ is satisfied, two of the four terms in $H^{ch}_{\perp}$ dominate, and tend to lock the combination of fields $2(\phi_{l}+\phi_{0})\pm l(\theta_{l}-\theta_0)$ to a fixed value, generating spontaneous current loops with opposite chiralities on the two inter-chain triangles (dashed circles in Fig. \ref{fig:ladders}). It is important to mention that the conventional (typically more relevant) inter-chain coupling term $\cos\left(\theta_l-\theta_{l'}\right)$, which in standard spin-ladders forces spins of adjacent chains to order in the $XY$-plane, is not present here because of the triangular structure frustration.

\section{Phase diagram}
\label{sec:PhaseDiagram}
\begin{figure}
	
	\begin{tikzpicture}[scale=1.7]

\draw[->] (0,0) -- (4.2,0) node[below] {$K$};
\draw[->][fill=red] (0,0) -- (0,2.7) node[left] {$B$};

\draw[scale=1, fill=violet!80] (0,0) rectangle (4,2.5)  {};

\draw[scale=1, fill=blue!80] (0,0.6)..controls (1,0.6) and (1.3,0.1)..(2, 0) to (0,0) to (0,0.6);

\draw[scale=1, fill=red!80] (0.7, 1.5)..controls (1.2, 1.4) and (1.4, 1.0) ..(2, 1.0) to
(2,1.0)..controls  (2.6, 1.0) and (2.8, 1.4)..(3.2, 1.5) to
(3.2, 1.5)..controls (2.8, 1.6) and (2.6, 2.0)..(2,2.0) to
(2, 2.0)..controls (1.4, 2.0) and (1.2, 1.6)..(0.7, 1.5);

\draw[scale=1, dashed, black] (0.7, 1.5) -- (0, 1.5) node[left] {$B_{D}$};
\draw[scale=1, dashed, black] (0.7, 1.5) -- (0.7, 0) {};
\draw[scale=1, dashed, black] (3.2, 1.5) -- (3.2, 0) {};

\node[circle, fill=black,inner sep=0pt,minimum size=2pt,] at (2,0) {};
\node[below] at (2,0) {$\frac{1}{2}$};
\node[circle, fill=black,inner sep=0pt,minimum size=2pt,] at (0.7, 0) {};
\node[below] at (0.7, 0) {$\frac{1}{2}-\delta$};
\node[circle, fill=black,inner sep=0pt,minimum size=2pt,] at (3.2, 0) {};
\node[below] at (3.2, 0) {$\frac{1}{2}+\delta$};

\node[scale=2, black] at (0.5, 0.25) {$VBC$};
\node[scale=2, black] at (2, 1.5) {$CSL$};
\node[scale=2, black] at (2.4, 0.5) {$MSL$};

\end{tikzpicture}
	\caption{(Colors online) Schematic phase diagram emanating from the low-energy Hamiltonian as a function of the Luttinger parameter $K$ and the magnetic field $B$. To obtain it we assume that $K_{0}\simeq K$; $B_D\propto D$ is the value of $B$ which exactly obeys the commensurability condition $k_B=\pm k_D/4$ (see text). Red color corresponds to a CSL phase, blue to a VBC and violet to a MSL.}
\label{fig:RelevanceOfCosine}
\end{figure}
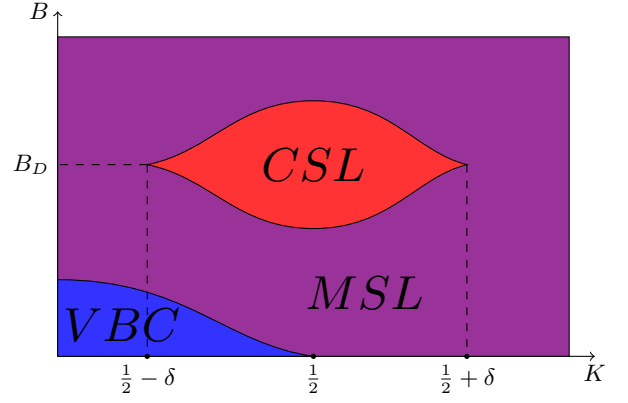

Having derived the low energy Hamiltonian [Eqs. (\ref{eq:H_LowEnergy}) through (\ref{eq:CommensurateCond})], we next obtain the phase diagram by employing perturbative renormalization group (RG) to analyze the effect of various terms (see the resulting diagram Fig. \ref{fig:RelevanceOfCosine}). In what follows, we regard the spin interaction parameters as fixed and consider the magnetic field $B$ as a tuning parameter.

For a general interaction term
\begin{align}
\frac{2g_{v}}{\left(2\pi a\right)^{2}}\cos\left(\lambda\phi+\tilde{\lambda}\theta\right)
\end{align}
added to a quadratic part in the form of a LL, the corresponding RG equations are
\begin{align}
\frac{\mathrm{d}K}{\mathrm{d}l}&=\frac{1}{16}\left[\tilde{\lambda}^{2}-\lambda^{2}K^{2}\left(l\right)\right]g_{v}^{2}\left(l\right) \label{eq:GeneralRG1} \\
\frac{\mathrm{d}g_{v}}{\mathrm{d}l}&=\left[2-\frac{1}{4}\left(\lambda^{2}K+\tilde{\lambda}^{2}\frac{1}{K}\right)\right]g_{v}\left(l\right).
\end{align}
Our model includes two interaction terms of this form [Eqs. (\ref{eq:H_cos}) and (\ref{eq:H^ch_perp})], however each is typically suppressed by a rapid oscillating factor. For low magnetic fields ($k_{l}^{B}\rightarrow0$) the cosine $H_l^{int}$ Eq. (\ref{eq:H_cos}) becomes relevant for $K<\frac{1}{2}$ and favors dimerization within each chain, where spins on adjacent sites form singlets. The choice of dimer configuration is arbitrary, which leads to a spontaneous symmetry breaking and formation of a VBC. Subsequently, spin excitations are gapped, making it a spin insulator. For stronger magnetic fields such that $k_{l}^{B}$ is significant, this phase melts via a commensurate-incommensurate type transition.

For typical values of the magnetic field $B$, the term $H^{ch}_{\perp}$ [Eq. (\ref{eq:H^ch_perp})] is suppressed as well for the same reason. However, tuning the ratio between $B$ and the DM coefficient $D$ to the commensurate value
\begin{align}
\nu_s \equiv k^B/k^D\simeq \pm\frac{1}{4}
\end{align}
yields $\Delta k_{\pm}\simeq0$ [Eq. (\ref{eq:CommensurateCond})] and reduces the rapid phase oscillations.  In that case, the corresponding RG equation for $g_{\perp}^{ch}$ is
\begin{align}
\frac{\mathrm{d}g_{\perp}^{ch}}{\mathrm{d}l}&=\left[2-\Delta_{ch}\right]g_{\perp}^{ch}\left(l\right)
\label{eq:RGforg_ch}
\end{align}
with
\begin{align}
\Delta_{ch}\simeq\left(\frac{1}{2K}+2K\right)-\frac{1+4K^{2}}{4}\frac{g_{\perp}^{z}}{u},
\end{align}
where we assume that $K_0\approx K$, $u_0\approx u$. The chiral term is hence relevant for $\frac{1}{2}-\delta <K<\frac{1}{2}+\delta$, where $\delta=\sqrt{\frac{1}{8}\frac{g_{\perp}^{z}}{u}}$. In addition, using Eq. (\ref{eq:GeneralRG1}) with $\lambda=2$ and $\tilde{\lambda}=1$, we note that when $g_{\perp}^{ch}$ flows to strong coupling, $K$ flows to the stable fixed point $K=\frac{\tilde{\lambda}}{\lambda}=\frac{1}{2}$. Namely, $SU(2)$ symmetry is recovered. To understand the nature of the order induced by this term it is natural to employ the chiral representation of bosonic fields
\begin{align}
\phi_{R,l}&=\frac{1}{2}\theta_{l}-\phi_{l}\\
\phi_{L,l}&=\frac{1}{2}\theta_{l}+\phi_{l}.
\end{align}
In the chiral basis $H_{\perp}^{ch}$ couples left(right)-movers to right(left)-movers in adjacent chains, leaving two counter-propagating modes on the outer chains. The resulting state is a chiral spin liquid (CSL) with a quantized Hall heat conductance. It follows from the analysis above that this phase is stable in the finite region in $K-B$ plane colored red in Fig. \ref{fig:RelevanceOfCosine}.

It should be emphasized, that the above estimated range of stability is based on a perturbative treatment of the term Eq. (\ref{eq:H^ch_perp}) which is capable of acquiring a vacuum expectation value, and provides a gap to excitations of the CSL. However, this is likely an underestimate of the robustness of the CSL phase. The chiral coupling constant $g_{\perp}^{ch}$ is further renormalized by additional terms in the Hamiltonian, which can not generate a mass by themselves but have a lower scaling dimension \cite{NLK93}. This point is further discussed in Sec. \ref{sec:sum_remarks}.

Finally, for intermediate, incommensurate values of $B$ where none of the cosine terms are relevant we are left with the quadratic part of the Hamiltonian, which is a gapless liquid we dub a metallic spin liquid (MSL) (the violet-colored region in Fig. \ref{fig:RelevanceOfCosine}). As shown in the next section, the distinction between the various phases is most prominently manifested by the behavior of their thermal Hall conduction.

\section{Thermal Hall Conductivity}
\label{sec:Thermal Hall Conductivity}
Now that we have identified the distinct phases dominating the system for different parameters, we turn to the calculation of thermal Hall conductivity characterizing each phase. The heat current operator along the strip direction is defined by the corresponding continuity equation:
\begin{align}
\partial_{t}\mathcal{H}\left(x\right)	=-\partial_{x}J_h\left(x\right),
\label{J_h_Def}
\end{align}
where $\mathcal{H}\left(x\right)$ is the energy density of  the  low-energy Hamiltonian [Eqs. (\ref{eq:H_LowEnergy})-(\ref{eq:H_cos}) and Eqs. (\ref{eq:H_perp})-(\ref{eq:H^ch_perp})]. To evaluate the left hand side of the above equation we assume the full Hamilton dynamics (including the terms $H_l^{int}$ and $H^{ch}_{\perp}$); details are given in App. B. The resulting operator has a quadratic form, and can be conveniently written as
\begin{align}
\label{eq:HeatCurrent}
& J_{h}\left(x\right)=\partial_{x}\Phi^{T}\hat{J}_{h}\partial_{x}\Phi \; ,\\ \nonumber
&\hat{J}_{h}\equiv\begin{pmatrix}Q_{1} & K_{1} & 0\\
K_{1}^{T} & Q_{0} & K_{-1}^{T}\\
0 & K_{-1} & Q_{-1}
\end{pmatrix}
\end{align}
where $\Phi^T =\begin{pmatrix}\theta_{1} & \phi_{1} & \theta_{0} & \phi_{0} & \theta_{-1} & \phi_{-1}\end{pmatrix}$,
\begin{align}
Q_{1}=\frac{u^{2}}{2\pi}\begin{pmatrix}2\alpha K; & 1+\alpha^{2}\\
1+\alpha^{2}; & 2\alpha\frac{1}{K}
\end{pmatrix}\\K_{1}=\frac{1}{2}\begin{pmatrix}g_{\perp}^{ch}u_{0}K_{0} & uKg_{\perp}^{z}\\
u_{0}K_{0}g_{\perp}^{z} & g_{\perp}^{ch}\frac{u}{K}
\end{pmatrix}
\end{align}
and
\begin{equation}
\alpha \equiv \frac{\pi g}{u}
\label{alpha_g_def}
\end{equation}
is a dimensionless parameter characterizing the chiral interaction in the chains [see Eq. (\ref{eq:H_LL})]; $K_{-1}$ and $Q_{-1}$ are obtained by taking $\alpha\rightarrow-\alpha$ and $g_{\perp}^{ch}\rightarrow -g_{\perp}^{ch}$, and $Q_0$ by the substitution $\alpha\rightarrow 0$, $u\rightarrow u_0$. It is worth pointing out that the cosine terms $H_{l}^{int}$ and $H^{ch}_{\perp}$, which are responsible for inducing the VBC and CSL phases, do not affect the form of $J_h$ (see App. B). However, in both strong coupling phases they prominently affect its expectation values.

To proceed with the calculation of the thermal Hall conductance, we introduce a thermal gradient across the strip assuming that the top and the bottom chains are held at temperatures $T_1$ and $T_{-1}$ respectively, where $T_{1/-1}=T\pm \frac{1}{2}\Delta T$ and $\Delta T\ll T$. The calculation of the resulting net heat current then follows a somewhat different path for each of the three phases, as described in detail below. However, in all cases it is dominated by contributions from two weakly coupled channels with opposite chiralities on the top and bottom sections of the Kagome strip, each approximately given by its local equilibrium value. This yields the linear response result $\langle J_{h}\rangle=\kappa_{xy}\Delta T$.

\subsection{Metallic Spin Liquid}
\label{sec:SimpleLiquid}
\begin{figure}
	
	\includegraphics[scale=0.35]{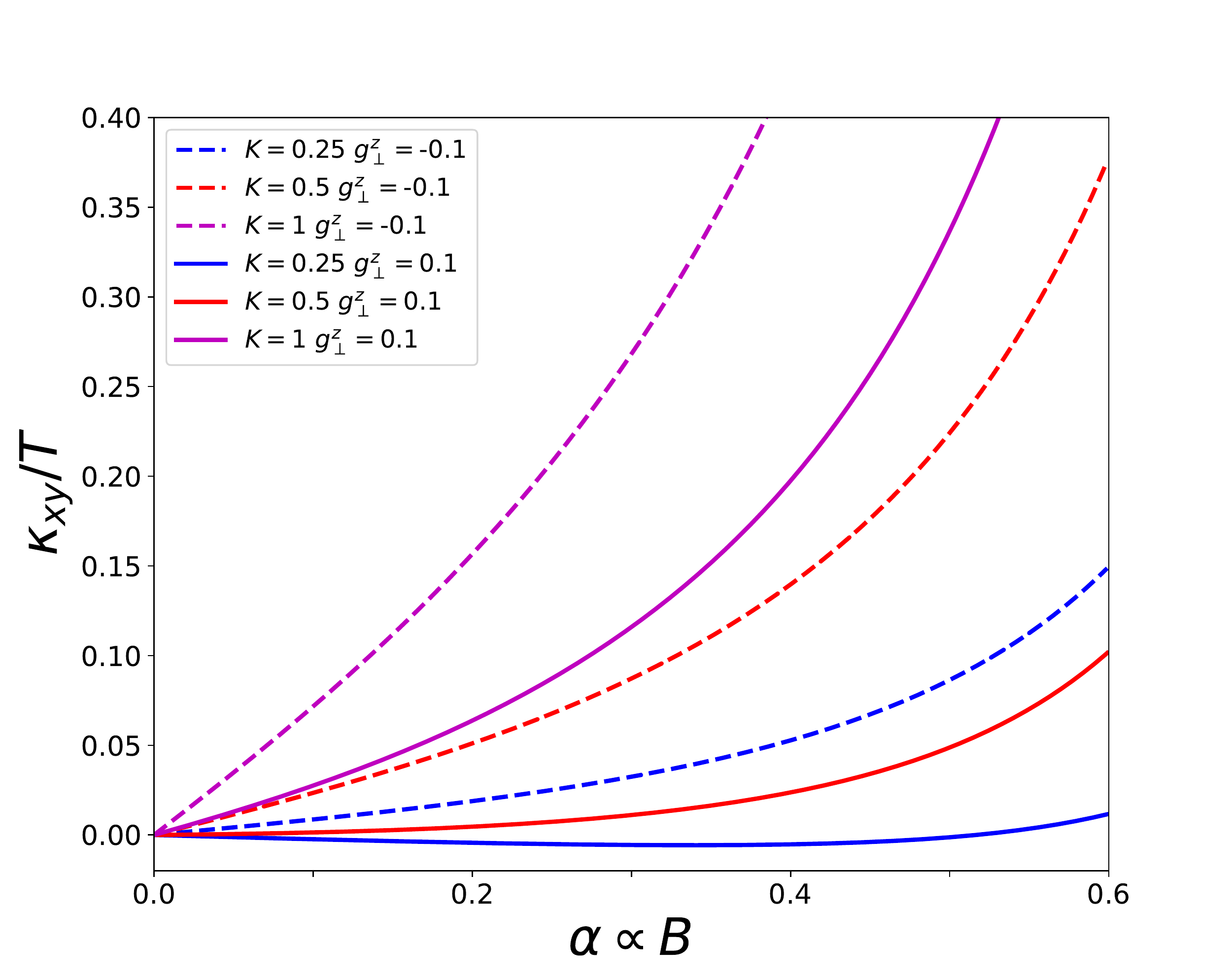}
	\caption{(Colors online) The thermal Hall conductance as a function of the magnetic field $B$ in the MSL phase, evaluated from Eq. (\ref{eq:HeatCurrentSL}) for different values of the Luttinger parameter $K$. Here $\frac{g^z_{\perp}}{u}\simeq \pm 0.1$, $\frac{g^{ch}_{\perp,0}}{uK}=0.03\alpha$, $u_0=1.1u$ and $K_0=1.1K$. Solid lines correspond to positive values of $g^{z}_{\perp}$ (AFM inter-chain interaction) and dashed to negative (FM interaction).}
\label{fig:deltaKappaXY}
\end{figure}

We first consider the MSL phase, established
when none of the cosine terms are relevant and we are left with the quadratic part of $H$. For convenience, we write the corresponding action and heat current operator [Eq. (\ref{eq:HeatCurrent})] in terms of chiral fields, defined via the transformation
\begin{align}
\label{eq:BasisMSL}
\Phi=\begin{pmatrix}A & 0 & 0\\
0 & U_{0} & 0\\
0 & 0 & A
\end{pmatrix}\Phi_{ch}
\end{align}
where $\Phi_{ch}^T=\begin{pmatrix}\phi_{1}^{R} & \phi_{1}^{L} & \phi_{0}^{R} & \phi_{0}^{L} & \phi_{-1}^{R} & \phi_{-1}^{L}\end{pmatrix}$ and
\begin{align}
\label{eq:Aand U0_def}
A=\frac{1}{2}\begin{pmatrix}\frac{1}{K} & \frac{1}{K}\\
-1 & 1
\end{pmatrix},\quad U_{0}=\frac{1}{2}\begin{pmatrix}\frac{1}{K_{0}} & \frac{1}{K_{0}}\\
-1 & 1
\end{pmatrix}.
\end{align}
In this basis, the action (at uniform $T$) acquires the form
\begin{align}
\label{eq:ActionMSL}
S&=\frac{T}{ 2L}\sum_{\vec{q}}\Phi_{ch}^{T}(-\vec{q})\begin{pmatrix}S_{1} & F_{1} & 0\\
F_{1}^{T} & S_{0} & F_{-1}^{T}\\
0 & F_{-1} & S_{-1}
\end{pmatrix}\Phi_{ch}(\vec{q})
\end{align}
where
\begin{align}
&S_{\pm1}=\frac{1}{2\pi K}\begin{pmatrix}q\left( u_{_\mp}q-i\omega_{n}\right)  & 0\\
0 & q\left( u_{_\pm}q+i\omega_{n}\right)
\end{pmatrix}\\&S_{0}=\frac{1}{2\pi K_{0}}\begin{pmatrix}q\left(u_{0}q-i\omega_{n}\right) & 0\\
0 & q\left(u_{0}q+i\omega_{n}\right)
\end{pmatrix}
\end{align}
and
\begin{align}
&F_{1}=\frac{q^{2}}{4}\begin{pmatrix}-\frac{g_{\perp,0}^{ch}}{K}+g_{\perp}^{z} & \frac{g_{\perp,0}^{ch}}{K}-g_{\perp}^{z}\\
-\frac{g_{\perp,0}^{ch}}{K}-g_{\perp}^{z} & \frac{g_{\perp,0}^{ch}}{K}+g_{\perp}^{z}
\end{pmatrix},\\&F_{-1}=F_{1}@\left(g_{\perp,0}^{ch}\rightarrow-g_{\perp,0}^{ch}\right).
\end{align}
Here $\Phi_{ch}(\vec{q})$ are the space-time Fourier components of the local field $\Phi_{ch}$, where $\vec{q}=(\omega_n,q)$ and $\omega_n$ are Matsubara frequencies; $u_{_\pm}\equiv\left(1\pm \alpha\right)u$.
The off-diagonal blocks are parametrized by the inter-chain interaction coefficients $g_{\perp,0}^{ch}$, $g_{\perp}^{z}$, which we treat perturbatively. In their absence (i.e. $g_{\perp,0}^{ch}=g_{\perp}^{z}=0$), heat flow is purely longitudinal and is carried by two counter-propagating modes at the {\em same} temperature on each side of the strip. As a result, even under application of a finite transversal thermal bias $\Delta T$, the net heat current $\langle J_{h}\rangle=0$.
To obtain the leading correction for finite inter-chain coupling we apply a perturbation expansion to second order in the coupling constants $g_{\perp,0}^{ch}$, $g_{\perp}^{z}$. Leaving the details of  the calculation to App. \ref{sec:ExpectationValues}, we get
\begin{align}
\label{eq:HeatCurrentSL}
&\langle J_{h}\rangle=\kappa_{xy}\Delta T \; ,\\ \nonumber
&\kappa_{xy}=TK_{0}K\frac{\pi}{3}\frac{\pi^{2}}{u_{0}^{2}}\Big[\frac{1}{K}g_{\perp,0}^{ch}g_{\perp}^{z}f_{s}\left(\alpha,\gamma\right)\\
&+\left(\left(\frac{g_{\perp,0}^{ch}}{K}\right)^{2}+\left(g_{\perp}^{z}\right)^{2}\right)f_{a}\left(\alpha,\gamma\right)\Big]\nonumber
\end{align}
where
\begin{align}
f_{a}\left(\alpha,\gamma\right)&=-\alpha\Big[\frac{1}{\left(1-\alpha^{2}\right)}+\frac{2\gamma}{\left(1-\alpha^{2}\right)^{2}}\\
&+\frac{\gamma\left(3\gamma-1\right)\left(\gamma+2\right)}{\left(\left(\gamma+1\right)^{2}-\alpha^{2}\right)\left(1-\alpha^{2}\right)}\Big]\\ \nonumber f_{s}\left(\alpha,\gamma\right)&=1+2\Big[\frac{1}{1-\alpha^{2}}+\frac{\gamma\left(1+\alpha^{2}\right)}{\left(1-\alpha^{2}\right)^{2}}\\
&+\frac{\gamma\left(3\gamma-1\right)\left(1+\gamma+\alpha^{2}\right)}{\left(\left(\gamma+1\right)^{2}-\alpha^{2}\right)\left(1-\alpha^{2}\right)}\Big] \nonumber
\end{align}
and $\gamma\equiv\frac{u_0}{u}$.
Recalling that $g_{\perp,0}^{ch}$ and $\alpha$ [proportional to $g$ via Eq. (\ref{alpha_g_def})] have the same origin (they are the coefficients of the time-reversal-breaking three-spin interactions), we assume them to be odd functions of the applied magnetic field $B$. Hence, for low $B$, as long as the MSL phase is stable the thermal Hall conductance is approximately linear:
\begin{align}
\kappa_{xy}\propto B\; .
\end{align}
Note, however, that the sign of the coefficient depends on details of the various parameters (see Fig. \ref{fig:deltaKappaXY}).

\subsection{Chiral Spin Liquid}
\label{sec:ChiralPhase}
A more remarkable behavior of $\kappa_{xy}$ is exhibited in the chiral CSL
phase, emerging in the vicinity of commensurate values of the magnetic field $B=\pm B_D$ (see Fig. \ref{fig:RelevanceOfCosine}).
As discussed in the previous section, in this phase the inter-chain chiral term $H^{ch}_{\perp}$ [Eq. (\ref{eq:H^ch_perp})] becomes relevant, and moreover renormalizes the Luttinger parameter to $K\rightarrow \frac{1}{2}$. Employing Eqs. (\ref{eq:BasisMSL}), (\ref{eq:Aand U0_def}) with $K=K_0=\frac{1}{2}$, the operators dominating $H_{\perp}^{ch}$ can be expressed in terms of the chiral fields in the following form \cite{chirality}:
\begin{align}
\label{eq:H_simple}
\mathcal{H}_{\perp}^{ch}\sim&\cos\left(2\phi^L_{0}-2\phi^R_{1}\right)+\cos\left(2\phi^L_{-1}-2\phi^R_{0}\right),
\end{align}
introducing two independent sine-Gordon models in the low-energy Hamiltonian. These operators acquire a vacuum expectation value, and generate a mass to fluctuations in the fields $(\phi^L_{0}-\phi^R_{1})$, $(\phi^L_{-1}-\phi^R_{0})$. As a consequence, we are left with two counter-propagating chiral modes on {\em opposite} edges of the strip, as $\phi^L_{1}$ and $\phi^R_{-1}$ remain gapless. In the presence of a thermal gradient $\Delta T=T_1-T_{-1}$, this leads to a quantized thermal Hall conductance
\begin{align}
\label{eq:HeatCurrentCL}
&\kappa_{xy}=\frac{\pi}{6}T,
\end{align}
which is exactly what we expect for one mode per edge to contribute. There are corrections to the quantized value resulting from the inter-chain coupling, but they are exponentially suppressed due to the bulk gap and thus negligible relatively to $\frac{\pi}{6}$. We note, however, that as $B$ deviates from the ideal value $\pm B_D$, the gap is suppressed approaching a commensurate-incommensurate transition. This enhances the deviation from the universal value and causes an overall reduction of $\kappa_{xy}$. We thus predict a plateau in $\kappa_{xy}$ vs. $B$ centered at $B=B_D$, as indicated in Fig. \ref{fig:IntroPlot}.

\subsection{Valence Bond Crystal}
\label{sec:DimerPhase}
We finally focus on the VBC phase dominating the low $B$, low $K$ regime where the interaction term (\ref{eq:H_cos}) is relevant and induces dimerization in each chain, resulting an ordered pattern of spin singlets \cite{Giamarchi}. The effective low-energy theory, describing fluctuations of the Bosonic fields $\phi_l$ around the favored values $\pm\pi/4$, is massive.
To calculate the heat current, we exploit the fact that deep in this phase there is a point of free massive fermions ($K=\frac{1}{4}$) for which we can treat the intra-chain terms exactly. The Hamiltonian $H_l$ acquires the form
\begin{align}
\label{eq:FermionHamiltonian}
H_{f}^{(l)}&=\int\mathrm{d}x\Big[u\left(\psi_{R}^{\dagger}(-i\partial_{x})\psi_{R}-\psi_{L}^{\dagger}(-i\partial_{x})\psi_{L}\right) \nonumber \\
&-E\left(\psi_{R}^{\dagger}\psi_{L}+\psi_{L}^{\dagger}\psi_{R}\right)\Big] \nonumber \\
&+\pi lg\int\mathrm{d}x\Big\{\psi_{R}^{\dagger}(-i\partial_{x})\psi_{R}+\psi_{L}^{\dagger}(-i\partial_{x})\psi_{L}\Big\}
\end{align}
where $E\sim g_l$ is the energy gap to excitations.
The heat current in the Fermionic representation is given by
\begin{equation}
\label{eq:HeatCurrent_f}
J_h\simeq \sum_l J_{f}^{(l)}
\end{equation}
where
\begin{align}
\label{eq:HeatCurrentDC}
J_{f}^{(l)}\left(x\right)&=u^{2}\left(1-l\alpha\right)^2\psi_{R}^{\dagger}(-i\partial_{x})\psi_{R} \nonumber \\
&u^{2}\left(1+l\alpha\right)^2\psi_{L}^{\dagger}(-i\partial_{x})\psi_{L}
\end{align}
and we neglect subdominant corrections due to inter-chain coupling.
In the presence of a thermal gradient, the expectation value of each term $J_{f}^{(l)}$ is evaluated at the corresponding local equilibrium temperature $T_l=T+\frac{l}{2}\Delta T$. Taking the large gap limit ($E\gg T$), this yields
\begin{align}
\label{eq:HeatCurrentDimerFinal}
\langle J_h\rangle &=\kappa_{xy}\Delta T,
\end{align}
where
\begin{align}
&\kappa_{xy}\simeq\frac{E^{\frac{3}{2}}}{T^{\frac{1}{2}}}e^{-\frac{E}{T}}\times f\left(\alpha\right) \nonumber
\end{align}
and
\begin{align}
f\left(\alpha\right)&=\frac{2\sqrt{2}}{\sqrt{\pi}}\alpha\left(1+\alpha^{2}\right) \nonumber
\end{align}
 (see App. \ref{sec:ExpectationValues} for details). Inter-chain interactions induce even smaller exponential corrections to $\kappa_{xy}$, which we therefore neglect. This activated suppression of $\kappa_{xy}$ dominates as long as $T$ is below the gap $E$; as the latter is maximized for $B\rightarrow 0$, we obtain the behavior depicted in the lower $B$ part of Fig. \ref{fig:IntroPlot}.

\section{summary and concluding remarks}
\label{sec:sum_remarks}
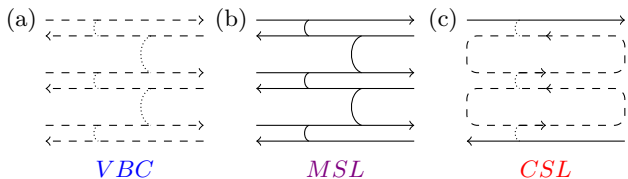
\begin{figure}
	
	\begin{tikzpicture}[scale=0.7]
	
	\coordinate (a) at (4,0);
	
	\node[scale=1, left] at (1,-1) {$(\mathrm{a})$};
	
	\draw[black, dashed, -{>[scale=2.0]}] (1,-1) -- (4,-1);
	\draw[black, dashed, -{>[scale=2.0]}] (4,-1.3) -- (1,-1.3);
	\draw[black, dashed, -{>[scale=2.0]}] (1,-2) -- (4,-2);
	\draw[black, dashed, -{>[scale=2.0]}] (4,-2.3) -- (1,-2.3);
	\draw[black, dashed, -{>[scale=2.0]}] (1,-3) -- (4,-3);
	\draw[black, dashed, -{>[scale=2.0]}] (4,-3.3) -- (1,-3.3);
	
	\node[blue, scale=1, below] at (2.5,-3.5) {$VBC$};
	
	\draw [black, densely dotted]   (2,-1) to[out=200,in=-200] (2,-1.3);
	\draw [black, densely dotted]   (3,-1.3) to[out=200,in=-200] (3,-2);
	\draw [black, densely dotted]   (2,-2) to[out=200,in=-200] (2,-2.3);
	\draw [black, densely dotted]   (3,-2.3) to[out=200,in=-200] (3,-3);
	\draw [black, densely dotted]   (2,-3) to[out=200,in=-200] (2,-3.3);
	
	\node[scale=1, left] at ($(1,-1)+(a)$) {$(\mathrm{b})$};
	
	\draw[black,  -{>[scale=2.0]}] ($(1,-1)+(a)$) -- ($(4,-1)+(a)$);
	\draw[black,  -{>[scale=2.0]}] ($(4,-1.3)+(a)$) -- ($(1,-1.3)+(a)$);
	\draw[black,  -{>[scale=2.0]}] ($(1,-2)+(a)$) -- ($(4,-2)+(a)$);
	\draw[black,  -{>[scale=2.0]}] ($(4,-2.3)+(a)$) -- ($(1,-2.3)+(a)$);
	\draw[black,  -{>[scale=2.0]}] ($(1,-3)+(a)$) -- ($(4,-3)+(a)$);
	\draw[black,  -{>[scale=2.0]}] ($(4,-3.3)+(a)$) -- ($(1,-3.3)+(a)$);
	
	\node[violet, scale=1, below] at ($(2.5,-3.5)+(a)$) {$MSL$};
	
	\draw [black]   ($(2,-1)+(a)$) to[out=200,in=-200] ($(2,-1.3)+(a)$);
	\draw [black]   ($(3,-1.3)+(a)$) to[out=200,in=-200] ($(3,-2)+(a)$);
	\draw [black]   ($(2,-2)+(a)$) to[out=200,in=-200] ($(2,-2.3)+(a)$);
	\draw [black]   ($(3,-2.3)+(a)$) to[out=200,in=-200] ($(3,-3)+(a)$);
	\draw [black]   ($(2,-3)+(a)$) to[out=200,in=-200] ($(2,-3.3)+(a)$);
	
	\node[scale=1, left] at ($(1,-1)+2*(a)$) {$(\mathrm{c})$};
	
	\draw[black, -{>[scale=2.0]}] ($(1,-1)+2*(a)$) -- ($(4,-1)+2*(a)$);
	\draw[black, dashed, rounded corners, -{>[scale=2.0]}] ($(2.5,-1.3)+2*(a)$) -- ($(1,-1.3)+2*(a)$)
	-- ($(1,-2)+2*(a)$) -- ($(4,-2)+2*(a)$) -- ($(4,-1.3)+2*(a)$)-- ($(2.5,-1.3)+2*(a)$);
	\draw[black, dashed, -{>[scale=2.0]}] ($(2,-2)+2*(a)$) -- ($(2.5,-2)+2*(a)$);
	\draw[black, dashed, rounded corners, -{>[scale=2.0]}] ($(2.5,-2.3)+2*(a)$) -- ($(1,-2.3)+2*(a)$)
	-- ($(1,-3)+2*(a)$) -- ($(4,-3)+2*(a)$) --($(4,-2.3)+2*(a)$)--($(2.5,-2.3)+2*(a)$);
	\draw[black, dashed, -{>[scale=2.0]}] ($(2,-3)+2*(a)$) -- ($(2.5,-3)+2*(a)$);

	\draw[black, -{>[scale=2.0]}] ($(4,-3.3)+2*(a)$) -- ($(1,-3.3)+2*(a)$);
	
	\node[red, scale=1, below] at ($(2.5,-3.5)+2*(a)$) {$CSL$};
	
	\draw [black, densely dotted]   ($(2,-1)+2*(a)$) to[out=200,in=-200] ($(2,-1.3)+2*(a)$);
	\draw [black, densely dotted]   ($(2,-2)+2*(a)$) to[out=200,in=-200] ($(2,-2.3)+2*(a)$);
	\draw [black, densely dotted]   ($(2,-3)+2*(a)$) to[out=200,in=-200] ($(2,-3.3)+2*(a)$);

	\end{tikzpicture}
	\caption{Sketch of the chiral modes pattern in each of the three phases. Here full lines represent gapless modes, dashed lines massive modes, and dotted lines denote coupling between the modes. (a) All the modes are gapped so that the heat current is exponentially suppressed. (b) The modes are all gapless, and the net heat current is resulting from the coupling between the chains. (c) Only one of the edge modes on each side is gapless, yielding a quantized $\kappa_{xy}$ of $\frac{\pi}{6}$. }
	\label{fig:SchematicPlotSum}
\end{figure}

In this paper we have studied a quasi-1D toy model for quantum spins with chiral interactions, focusing on a strip of the distorted Kagome lattice structure depicted in Fig. \ref{fig:ladders}. We
showed that this system possesses three distinct phases (see Fig. \ref{fig:RelevanceOfCosine}), stabilized in different regions of a parameter space including a Luttinger parameter $K$ (parametrizing the $XXZ$-anisotropy of spin exchange interactions) and a tunable magnetic field $B$. In the low $B$ regime, the spins form a VBC with gapped counter-propagating modes on the opposite edges of the strip; it therefore exhibits a `spin-insulator'-like exponential suppression of the thermal Hall conductance $\kappa_{xy}$ at low $T$ (and similarly of the longitudinal thermal conductivity, which we did not explicitly calculate). As the magnetic field is increased, destroying the singlet-crystal order of VBC, a `metallic' spin liquid (MSL) phase emerges, characterized by the thermal Hall conductance being linear in $T$ with a non-universal coefficient. By further increase in $B$, it reaches the vicinity of a commensurate value of $B_D$ favoring the formation of a CSL with a plateau of $\kappa_{xy}/T$ at $\frac{\pi}{6}$, resulting from the effective decoupling of two counter-propagating edge modes on opposite sides of the strip (see Fig. \ref{fig:SchematicPlotSum}). The transitions from one phase to another are transparently manifested in the behavior of the thermal Hall effect as a function of $B$ (see Fig. \ref{fig:IntroPlot}).

Among the three phases, the most intriguing is the CSL which exhibits a topological order. Notably, in our model it is restricted to a narrow range of $B$ surrounding a ``magic" value $B_D\propto D$; here $D$ denotes the strength of a DM interaction, which introduces a fictitious ``magnetic flux" due to the formation of spin-current loops within triangular plaquettes. The CSL therefore reflects a remarkable reminiscence to a FQH state in 2D charge conductors subject to a perpendicular magnetic field: with respect to the spinons, the magnetic field serves as a gate potential dictating their density compared to the particle-hole symmetric point $B=0$; a FQH liquid state is then established when this density is commensurate with the effective flux density proportional to $D$ or $-D$ (yielding a particle or hole-like FQH state, respectively). We emphasize, however, that the parameter $D$ (whose chirality can be traced back to spin-orbit interaction in the underlying material) is not analogous to a {\em uniform} magnetic field in an electronic system. Rather, in our model where we have introduced a distortion of the Kagome lattice with an explicit breaking of inversion symmetry in the transverse direction (see Fig. \ref{fig:ladders}), it induces flux of opposite sign on the top and bottom chains; i.e., on the two chains containing an {\em odd} number of triangles. This inversion symmetry breaking of the Star-of-David building-block is essential to the formation of the CSL phase.

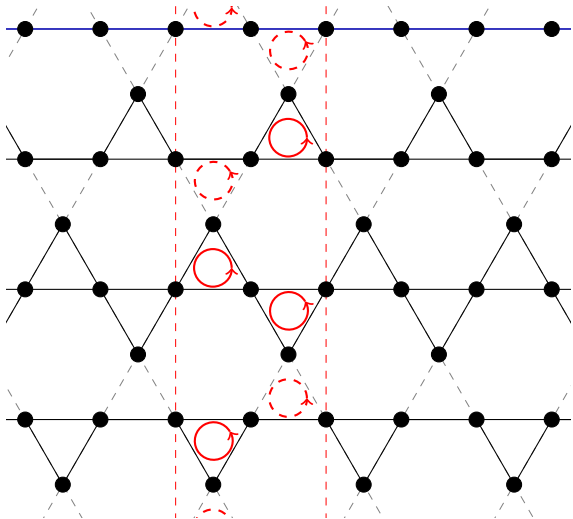
\begin{figure}
	
	\begin{tikzpicture}[scale=1]
	\clip (-1.25,-1.30) rectangle (6.25,5.5);
	\coordinate (a1) at (2,0);
	\coordinate (a2) at (1,{sqrt(3)});	
	\coordinate (d1) at (0,0);
	\coordinate (d2) at (1,0);	
	\coordinate (d3) at (0.5,{0.5*sqrt(3)});
	\coordinate (c) at ($0.33*(d1)+0.33*(d2)+0.33*(d3)$);
	\draw [style=help lines,dashed,shorten >=-10cm,shorten <=-10cm,red] ($0.5*(a1)$) -- ($2*(a2)-0.5*(a1)$) {};
	\draw [style=help lines,dashed,shorten >=-10cm,shorten <=-10cm,red] ($0.5*(a1)+1*(a1)$) -- ($2*(a2)+0.5*(a1)$) {};
	
	\foreach \n in {-3,-2,...,3}{
		\draw [shorten >=-10cm,shorten <=-10cm] ($\n*(a2)$) -- ($(a1)+\n*(a2)$);
		\draw [style=help lines,dashed,shorten >=-10cm,shorten <=-10cm] ($\n*(a1)$) -- ($(a2)+\n*(a1)$) {};
		\draw [style=help lines,dashed,shorten >=-10cm,shorten <=-10cm] ($0.5*(a1)+0.5*\n*(a2)+0.5*\n*(a1)$) -- ($0.5*(a2)+0.5*\n*(a2)+0.5*\n*(a1)$) {};
	}
	\draw[blue, shorten >=-10cm,shorten <=-10cm] ($3*(a2)$) -- ($3*(a2)+(a1)$) ;
	
	\foreach \n in {-3,-2,...,4}{
		\foreach \m in {-3,-2,...,4}{
			\node[draw,circle,inner sep=2pt,fill] at ($\n*(a1)+\m*(a2)+(d1)$) {};
			\node[draw,circle,inner sep=2pt,fill] at ($\n*(a1)+\m*(a2)+(d2)$) {};
			\node[draw,circle,inner sep=2pt,fill] at ($\n*(a1)+\m*(a2)+(d3)$) {};
		}
	}
	
	\draw ($(0,0)-(a1)$) -- ($(0,0)-0.5*(a1)$) -- ($(0,0)-0.5*(a2)$) -- (0,0) -- ($0.5*(a1)$) -- ($(a1)-0.5*(a2)$) -- ($(a1)$) -- ($0.5*(a1)+(a1)$) -- ($(a1)+(a1)-0.5*(a2)$) -- ($(a1)+(a1)$) -- ($0.5*(a1)+2*(a1)$) -- ($3*(a1)-0.5*(a2)$) -- ($3*(a1)$)-- ($3.5*(a1)$) ;
	
	\draw ($0.5*(a2)-(a1)$) -- ($(a2)-(a1)$) -- ($(a2)+0.5*(a2)-(a1)$) -- ($(a2)-0.5*(a1)$) -- ($0.5*(a2)$) -- ($(a2)$) -- ($(a2)+0.5*(a2)$) -- ($(a2)+0.5*(a1)$) -- ($0.5*(a2)+(a1)$) -- ($(a2)+(a1)$) -- ($(a2)+0.5*(a2)+(a1)$) -- ($(a2)+1.5*(a1)$) -- ($0.5*(a2)+2*(a1)$) -- ($(a2)+2*(a1)$) -- ($(a2)+0.5*(a2)+2*(a1)$) -- ($(a2)+2.5*(a1)$) -- ($0.5*(a2)+3*(a1)$);

	\draw ($2.5*(a2)-2*(a1)$) -- ($2*(a2)-1.5*(a1)$) -- ($2*(a2)-(a1)$) -- ($2.5*(a2)-(a1)$) -- ($2*(a2)-0.5*(a1)$) -- ($2*(a2)$) -- ($2.5*(a2)$) -- ($2*(a2)+0.5*(a1)$) -- ($2*(a2)+(a1)$) -- ($2.5*(a2)+(a1)$) -- ($2*(a2)+1.5*(a1)$) -- ($2*(a2)+2*(a1)$) -- ($2.5*(a2)+2*(a1)$) -- ($2*(a2)+2.5*(a1)$)  ;
	\foreach \n in {1}{        		
		\foreach \m in {0}{
			\draw[thick,dashed,red,-{>[scale=2.0]}] ([shift=(0:0.25cm)]$\n*(a1)+\m*(a2)+(c)$) arc (0:360:0.25cm);
			\draw[thick,red,-{>[scale=2.0]}] ([shift=(30:0.25cm)]$\n*(a1)+\m*(a2)-(c)$)arc (30:390:0.25cm);
		}
	}
	\draw[thick,red,-{>[scale=2.0]}] ([shift=(0:0.25cm)]$0*(a1)+1*(a2)+(c)$) arc (0:360:0.25cm);
	\foreach \n in {0}{        		
		\foreach \m in {2}{
			\draw[thick,red,-{>[scale=2.0]}] ([shift=(0:0.25cm)]$\n*(a1)+\m*(a2)+(c)$) arc (0:360:0.25cm);
			\draw[thick,dashed,red,-{>[scale=2.0]}] ([shift=(30:0.25cm)]$\n*(a1)+\m*(a2)-(c)$)arc (30:390:0.25cm);
		}
	}
	\draw[thick,red,-{>[scale=2.0]}] ([shift=(30:0.25cm)]$1*(a1)+1*(a2)-(c)$)arc (30:390:0.25cm);
	
	
	\draw[thick,dashed,red,-{>[scale=2.0]}] ([shift=(0:0.25cm)] $3*(a2)-1*(a1)+(c)$) arc (0:360:0.25cm);
	\draw[thick,dashed,red,-{>[scale=2.0]}] ([shift=(30:0.25cm)] $3*(a2)-0*(a1)-(c)$) arc (30:390:0.25cm);
	\draw[thick,dashed,red,-{>[scale=2.0]}] ([shift=(0:0.25cm)] $1*(a1)-1*(a2)+(c)$) arc (0:360:0.25cm);
	\end{tikzpicture}
	\caption{(Colors online). An extension of the Kagome strip where we add an additional chain to the top edge (solid blue line). Duplicating this unit in the vertical direction yields a periodic 2D structure.}
\label{fig:2DExtension}
\end{figure}

The above described behavior can persist into a fully 2D Kagome lattice, provided it undergoes the appropriate distortion. A possible extension of our model to a 2D periodic structure can be constructed by adding a simple (linear) $XXZ$-chain along one of the edges (see Fig. \ref{fig:2DExtension}). The resulting pattern can then be duplicated to a periodic lattice in the transverse direction, with unit cell consisting of four weakly-coupled chains: two of them contain an odd number of triangles per (longitudinal) unit cell, supporting a well-defined chirality of spin-current, and two non-chiral ones containing an even number (2 or 0). In such a 2D structure under a suitable choice of parameters, a 2D CSL phase can form where the bulk is gapped, and only the outer chiral modes contribute to the thermal Hall conductance, thus maintaining the phenomenology of CSL. While our model is artificial in the sense that it assumes a particular generalization of the ideal Kagome lattice, we argue that qualitatively similar ingredients might play a role in other realizations of a CSL state, and manifest themselves in the observation of a quantized plateau in $\kappa_{xy}$ vs. $B$ as in Ref. \onlinecite{Kasahara2018}.

We finally comment on the possible effect of additional contributions to the model Hamiltonian allowed by symmetry, which we did not account for in our study. First, similarly to the FQHE, other commensurate ratios of $B$ and $D$ besides $B\simeq B_D$ may favor additional CSL states. The operators supporting the formation of such states are of the general form
\begin{align}
\cos\left(\theta_{0}-\theta_{l}+2n\phi_{l}+2n\phi_{0}-k_{D}x-4nk_{B}x\right)
\end{align}
with an arbitrary integer $n>1$. However, except for $n=1$ such terms are typically irrelevant. More interesting is the effect of additional terms arising from the chiral interactions, which can not acquire a vacuum expectation value but significantly contribute to the flow of the chiral coupling constant $g^{ch}_{\perp}$ under RG. This includes operators of the form $\sim\sin(\theta_{\pm 1}-\theta_0\mp k_Dx)$ which are frustrated due to the finite value assumed for $k_D\propto D$, as well as chiral operators of the form $\sim\cos(2\phi_{\pm 1}\pm \theta_{\pm 1}\mp\theta_0)$. While they can not generate a mass, their relatively low scaling dimension dictates important corrections to the RG equation Eq. (\ref{eq:RGforg_ch}) which drive $g^{ch}_{\perp}$ to strong coupling in a wider range of parameters \cite{NLK93}.

The above described operators may assist in maintaining the robustness of the CSL phase against disorder, e.g. due to defects in the perfect lattice structure, which inevitably exists to some degree in any realistic system and typically raises a serious concern in 1D systems. Its effect may be introduced via random variations of the exchange coupling constants
\begin{align}
J\rightarrow J+\delta J\left(x\right).
\end{align}
In the presence of a finite field $B$, this introduces coupling to the backscattering operator $\cos(2\phi_l)$ in each chain $l$. Such term is obviously more relevant than the interaction terms inducing the interesting phases in the clean limit for a wide range of the Luttinger parameter $K$. In particular, for $K<3/2$, the disorder is relevant and tends to induces localization of the spin excitations \cite{GianarchiShulz1988} in the limit $T\rightarrow 0$. Recalling that AFM spin chains (where $K<1$) are fully included in this regime, this appears to severely challenge the possibility to observe a CSL behavior. However, competition with relevant chiral terms can shift the disorder-dominated localized phase to lower values of $K$. Either way, disorder poses a practical limitation on the observation of CSL in real materials: similarly to FQH states in electronic systems, the samples have to be sufficiently clean that the characteristic energy scale ($\Delta_{dis}$) associated with disorder, the energy gap ($\Delta_{csl}$) to excitations of the CSL and the temperature $T$ of the measurement obey the hierarchy $\Delta_{dis}\ll T\ll \Delta_{csl}$. Lastly, we note that since $K$ is an arbitrary parameter in our theory, it can be readily extended to spin systems with ferromagnetic (FM) interactions ($K>1$) which are more immune to disorder. Indeed, the thermal Hall measurement of Ref. \onlinecite{HirschbergerChinellLeeOng2015} was performed on a FM Kagome compound \cite{OferMarciparChandraGazitPodolskyArovasKeren}; rather than a CSL, the data indicate a behavior qualitatively consistent with the MSL phase dominating the high $K$ region of Fig. \ref{fig:RelevanceOfCosine}. In this regime, though, a spin-wave theory  \cite{HyunyongJungPatrick} provides a more suitable approach to the 2D system.

\acknowledgements
Useful discussions with Sam Carr, Eyal Leviatan, David Mross, Raul Santos, Eran Sela and Chandra Varma
are gratefully acknowledged. P. T. thanks the Bar-Ilan Institute for Nanotechnology and Advanced Materials for financial support during the academic year 2017. E. S. thanks the Aspen Center for Physics (NSF Grant No. 1066293) for its hospitality. This
work was supported by the US-Israel Binational Science Foundation
(BSF) grant 2016130, and the Israel Science Foundation (ISF)
grant 231/14.

\appendix

\section{Bosonizationon}
\label{sec:Bosonizationon}
In this Appendix we present some details of the derivation of a bosonized form for several non-standard terms in the Hamiltonian Eq. (\ref{eq:H_original}), particularly including the chiral terms. Throughout the derivation we keep only the most relevant operators. It is important to point out that we perform a "shift" transformation [Eq. (\ref{eq:shift_trans})]  to eliminate linear terms that are induced by the Bosonization, but it does not affect quadratic terms and results in oscillation for the cosine (sine) terms. We comment about these oscillations where appropriate.

We start with the exchange terms $\sum_{\langle i,j \rangle}S^{+}_{l,i}S^{-}_{l,j}$,  and employ a JW transformation for each of the chains:
\begin{align}
\label{eq:JordanWignerSz}
S_{l,i}^{z}&=C_{l,i}^{\dagger}C_{l,i}-\frac{1}{2} \\
\label{eq:JordanWignerSplus}
S_{l,i}^{+}&=C_{l,i}^{\dagger}\left(\frac{1}{2}e^{i\pi\sum_{j<i}C_{l,j}^{\dagger}C_{l,j}}+h.c.\right),
\end{align}
where $C_{l,i}^{\dagger}$ and $C_{l,i}$ are spinless Fermions. For the lower chain $l=-1$ (see Fig. \ref{fig:ladders}), this yields
\begin{align}
&\sum_{\langle i,j\rangle} S_{-1,i}^{+}S_{-1,j}^{-} =  \nonumber \\
&\sum_{n}\Big\{ S_{3n}^{+}S_{3n+1}^{-}+S_{3n+1}^{+}S_{3n+2}^{-}+S_{3n}^{-}S_{3n+2}^{+} +S_{3n+2}^{+}S_{3n+3}^{-}\Big\}\nonumber \\
&=\sum_{n}\sum_{m=0,1,2}\Big\{C_{3n+m}^{\dagger}C_{3n+m+1}\\
&+\sum_{n}C_{3n}^{\dagger}\left(1-2C_{3n+1}^{\dagger}C_{3n+1}\right)C_{3n+2}\Big\}  \nonumber \\
&=-\sum_{n}\sum_{m=0,1,2}\Big\{C_{3n+m}^{\dagger}C_{3n+m+1} \nonumber \\
& -2\sum_{n}C_{3n}^{\dagger}C_{3n+2}S_{3n+1}^{z}\Big\}\; , \nonumber
\end{align}
where in the last step we performed a transformation $C_{n}\rightarrow\left(-\right)^{n}C_{n}$. Bosonizing the fermions by
\begin{align}
\label{eq:Bosonization1}
C_{l,i}&\propto e^{ik_{F}x}\psi_{l,R}\left(x\right)+e^{-ik_{F}x}\psi_{l,L}\left(x\right)\\
\label{eq:Bosonization2}
\psi_{l,r}\left(x\right)&=\lim_{a\rightarrow0}\frac{1}{\sqrt{2\pi a}}e^{-i\left(r\tilde{\phi}_{l}\left(x\right)-\tilde{\theta}_{l}\left(x\right)\right)}
\end{align}
results in
\begin{align}
&\sum_{\langle i,j\rangle} \Big\{S_{-1,i}^{+}S_{-1,j}^{-}+h.c.\Big\}\nonumber \\ \label{eq:cosine_xy}
&\simeq\int\mathrm{d}x\frac{3}{\pi a^{2}}\left\{ \left(a\partial_{x}\tilde{\phi}_{-1}\right)^{2}+\left(a\partial_{x}\tilde{\theta}_{-1}\right)^{2}\right\} \\
&+\int\mathrm{d}x\frac{4}{\left(\pi a\right)^{2}}\left(a\partial_{x}\tilde{\phi}_{-1}\right)^{2}+\int\mathrm{d}x\frac{2}{\left(\pi a\right)^{2}}\cos4\tilde{\phi}_{-1} \; .  \nonumber
\end{align}
By symmetry, the continuum limit for the upper chain ($l=1$) is the same.
A similar calculation for the middle chain ($l=0$) gives a result that is only different by numerical factors. In particular, the coefficient of the last (cosine) term is doubled due to the presence of two triangles in the unit cell. Transformation to the "shifted" fields [Eq. (\ref{eq:shift_trans})] induces oscillation in the cosine term, which we discuss in the main text.

The Bosonization of $\sum_{\langle i,j \rangle}S^{z}_{l,i}S^{z}_{l,j}$  is simpler because we can use the closed form of $S^z$ [Eq. (\ref{eq:BosonizationMapping})] in terms of the corresponding Boson fields, leading to
\begin{align}
&\sum_{\langle i,j\rangle} S_{-1,i}^{z}S_{-1,j}^{z}=\sum_{n}\Big\{ S_{3n}^{z}S_{3n+1}^{z}+S_{3n+1}^{z}S_{3n+2}^{z} \nonumber \\
&+S_{3n}^{z}S_{3n+2}^{z}+S_{3n+2}^{z}S_{3n+3}^{z} \Big\} \label{eq:cosine_zz} \\
&\simeq\int\mathrm{d}x\Big\{\frac{6}{\pi^{2}}\left(\partial_{x}\tilde{\phi}_{-1}\right)^{2}-\frac{1}{\pi^{2}a^{2}}\cos4\tilde{\phi}_{-1}\Big\} \; . \nonumber
\end{align}
Again, for the upper chain we get the same result, and the middle chain result only differs by the prefactors of bosonic operators. Combining Eqs. (\ref{eq:cosine_xy}) and (\ref{eq:cosine_zz}), and accounting for the prefactors $\frac{1}{2}J_{\parallel}^{xy}$, $J_{\parallel}^{z}$ of the corresponding exchange terms, we find the overall coefficient of the $\cos(4\tilde{\phi}_{\pm 1})$ term to be $g_{\pm 1}\propto (J_{\parallel}^{xy}-J_{\parallel}^{z})$. Similarly, the coefficient of the $\cos(4\tilde{\phi}_{0})$ term is $g_{0}\propto (2J_{\parallel}^{xy}-J_{\parallel}^{z})$.

We now turn our attention to the last intra-chain interaction term, introducing the chiral operators $\vec{S}_{l,i}\cdot\left(\vec{S}_{l,j}\times\vec{S}_{l,k}\right)$. For the lower-most triangle (residing on $l=-1$), we obtain
\begin{align}
&\sum_{\left\{ i,j,k\right\} }\vec{S}_{i}\cdot\left(\vec{S}_{j}\times\vec{S}_{k}\right)=\sum_{n}\vec{S}_{3n}\cdot\left(\vec{S}_{3n+1}\times\vec{S}_{3n+2}\right) \nonumber \\
&=\sum_{n}\Big\{-i\left[C_{3n}^{\dagger}C_{3n+1}-C_{3n+1}^{\dagger}C_{3n}\right]S_{3n+2}^{z}  \nonumber\\
&-i\left[C_{3n+1}^{\dagger}C_{3n+2}-C_{3n+2}^{\dagger}C_{3n+1}\right]S_{3n}^{z} \\
&+2i\left[C_{3n}^{\dagger}C_{3n+2}-C_{3n+2}^{\dagger}C_{3n}\right]S_{3n+1}^{z}\Big\}  \nonumber \\
&\simeq \left( -i\frac{1}{2\pi a}4ia\partial_{x}\tilde{\theta}_{-1}\times2+\frac{4a}{\pi}\partial_{x}\tilde{\phi}_{-1}\partial_{x}\tilde{\theta}_{-1} \right) \left(-\frac{1}{\pi}\partial_{x}\tilde{\phi}_{-1} \right) \nonumber \\
&\simeq-\int\mathrm{d}x\frac{4}{\pi^{2}}\partial_{x}\tilde{\theta}_{-1}\partial_{x}\tilde{\phi}_{-1}\; .  \nonumber
\end{align}
For the upper chain ($l=1$) the above term gives the same result but with a {\em plus} sign, because of the opposite chirality on the top and bottom triangles. However, in the middle chain ($l=0$) this leading contribution cancels altogether, having contributions from triangles of both chiralities. The resulting low energy limit of the $3$-spin operator within $l=0$ is irrelevant, and hence neglected.

Next, we consider the inter-chain interactions. Here we present the coupling between the upper ($l=1$) and the middle ($l=0$) chains; the coupling between the bottom ($l=-1$) and middle ($l=0$) chains can then be inferred by symmetry.
Recalling the bosonic representation of the $S^{\pm}$ operator Eq. (\ref{eq:BosonizationMapping}), we get for the $xy$-exchange term
\begin{align}
&\sum_{\left\langle i,j\right\rangle }\left(S_{1,i}^{+}S_{0,j}^{-}+h.c.\right)=\\
&=\sum_{n}\Big\{S_{1,3n}^{+}S_{0,4n+1}^{-}+S_{1,3n+1}^{+}S_{0,4n+1}^{-}+h.c.\Big\}\nonumber\\
&\simeq\int\mathrm{d}x\Big\{\frac{1}{2\pi a}\cos\left(\tilde{\theta}_{1}\left(x\right)-\tilde{\theta}_{0}\left(x\right)\right)\nonumber\\
&\times\left[\left(-\right)^{3n}+\cos2\tilde{\phi}_{1}\left(x\right)\right]\nonumber\\
&+\frac{1}{2\pi a}\cos\left(\tilde{\theta}_{1}\left(x+a\right)-\tilde{\theta}_{0}\left(x\right)\right)\nonumber\\
&\times\left[\left(-\right)^{3n+1}+\cos2\tilde{\phi}_{1}\left(x+a\right)\right]\Big\}\nonumber\\
&\times\left[\left(-\right)^{4n+1}+\cos2\tilde{\phi}_{0}\left(x\right)\right]\; . \nonumber
\end{align}
Due to the staggering factor $\left(-\right)^{3n}$, the continuum limit of this term is dominated by operators of the form $\cos(\tilde{\theta}_{1}-\tilde{\theta}_{0})\cos 2\tilde{\phi}_{1}$ and $\cos(\tilde{\theta}_{1}-\tilde{\theta}_{0})\cos 2\tilde{\phi}_{1}\cos 2\tilde{\phi}_0$. The latter contributes to $H_{\perp}^{ch}$ [Eq. (\ref{eq:H^ch_perp})], and the former is a sum of chiral operators that can not generate a mass.
After performing the shift transformation [Eq. (\ref{eq:shift_trans})], one observes that these terms typically exhibit rapid phase oscillations.

For the $z$-component of the exchange coupling between chains $l=1$ and $l=0$, using Eq. (\ref{eq:BosonizationMapping}) we obtain a quadratic perturbation:
\begin{align}
&\sum_{\left\langle i,j\right\rangle }S_{1,i}^{z}S_{0,j}^{z}=\sum_{n}\Big\{\left(S_{1,3n}^{z}+S_{1,3n+1}^{z}\right)S_{0,4n+1}^{z} \nonumber \\
&\simeq\int\mathrm{d}x\frac{2}{\pi^{2}}\partial_{x}\tilde{\phi}_{1}\partial_{x}\tilde{\phi}_{0}.
\end{align}
The coupling between the middle ($l=0$) and the lower ($l=-1$) chains gives the same result, so that the terms are a part of $H_{\perp}^{0}$ [Eq.(\ref{eq:H^0_perp})].

Finally, we consider the chiral 3-spin operator
\begin{align}
&\sum_{\left\{ i,j,k\right\} }\vec{S}_{0,i}\cdot\left(\vec{S}_{1,j}\times\vec{S}_{1,k}\right)=\vec{S}_{0,4n+1}\cdot\left(\vec{S}_{1,3n+1}\times\vec{S}_{1,3n}\right) \nonumber \\
&=\sum_{n}\frac{i}{2}S_{0,4n+1}^{z}\cdot\left(S_{1,3n+1}^{+}S_{1,3n}^{-}-S_{1,3n+1}^{-}S_{1,3n}^{+}\right)  \\
&+\sum_{n}\frac{i}{2}S_{1,3n}^{z}\cdot\left(S_{0,4n+1}^{+}S_{1,3n+1}^{-}-S_{0,4n+1}^{-}S_{1,3n+1}^{+}\right) \nonumber \\
&+\sum_{n}\frac{i}{2}S_{1,3n+1}^{z}\cdot\left(S_{1,3n}^{+}S_{0,4n+1}^{-}-S_{1,3n}^{-}S_{0,4n+1}^{+}\right). \nonumber
\end{align}
Here, the first term has a simple form in the fermionic language
\begin{align}
\sum_{n}\frac{1}{2i}S_{0,4n+1}^{z}\cdot\left(C_{1,3n+1}^{\dagger}C_{1,3n}-C_{1,3n+1}C_{1,3n}^{\dagger}\right)
\end{align}
and induces gradient couplings between the outer chains and the middle one as a part of $H^{0}_{\perp}$ [Eq.(\ref{eq:H^0_perp})]. The last two terms generate many operators, most of which exhibit rapid oscillations. Among them we maintain the ones which contain oscillating phase factors depending on {\em both} wave-vectors $k_B$ and $k_D$, which can therefore cancel upon tuning them to a particular commensurate ratio. In particular, the following contribution couples to operators that are capable of acquiring a vacuum expectation value and induce the CSL phase:
\begin{align}
&\int\mathrm{d}x\Big\{\frac{1}{2\left(\pi a\right)^{2}}\cos\left(2\phi_{1}-2k_{1}^{B}x\right)\cos\left(2\phi_{0}-2k_{0}^{B}x\right) \nonumber \\
&\times\sin\left(\theta_{0}-k_{0}^{D}x-\theta_{1}+k_{1}^{D}\left(x+a\right)\right)\\
&+\frac{1}{2\left(\pi a\right)^{2}}\cos\left(2\phi_{1}-2k_{1}^{B}\left(x+a\right)\right) \nonumber \\
&\times\sin\left(\theta_{1}-k_{1}^{D}x-\theta_{0}+k_{0}^{D}x\right)\cos\left(2\phi_{0}-2k_{0}^{B}x\right)\Big\}. \nonumber
\end{align}
After applying trigonometric identities, this leads to an expression of the form $H^{ch}_{\perp}$ [Eq.(\ref{eq:H^ch_perp})]. 

\section{Derivation of the heat current operator}
\label{sec:HeatCurrentDerivation}
In this Appendix we present some details for the derivation of  the heat current density operator, which follows from the definition Eq. (\ref{J_h_Def}). The most common contribution to the left hand side, i.e. the commutator $i[H,\mathcal{H}]$, is arising from the LL Hamiltonian density $\mathcal{H}^{LL}$
\begin{align}
&i\int\mathrm{d}x'\left[\mathcal{H}^{LL}\left(x'\right),\mathcal{H}^{LL}\left(x\right)\right]=iu^{2}\pi^{2}\frac{1}{\left(2\pi\right)^{2}}\\
&\times\int\mathrm{d}x'\Bigg\{\left[\left(\partial_{x'}\phi\right)^{2},\Pi^{2}\left(x\right)\right]+\left[\Pi^{2}\left(x'\right),\left(\partial_{x}\phi\right)^{2}\right]\Bigg\}, \nonumber
\end{align}
which gives a well known result
\begin{align}
u^{2}\partial_{x}\left(\partial_{x}\phi\Pi\left(x\right)\right);
\end{align}
here and through out this section $\Pi\left(x\right)=\frac{1}{\pi}\partial_{x}\theta$. This result can be readily interpreted as a contribution ($J_h^{LL}$) to the right hand side of Eq. (\ref{J_h_Def}) where $J_h^{LL}=u^{2}\partial_{x}\phi\Pi\left(x\right)$. We proceed with a characteristic term included in the quadratic part of $H$, which is a commutation relation between $\mathcal{H}^{LL}$ and a gradient coupling of two fields:
\begin{align}
&i\int\left[\mathcal{H}^{LL}\left(x'\right),\partial_{x}\phi\partial_{x}\phi_{0}\right]=\\
&i\frac{u}{2\pi}\int\mathrm{d}x'\left[K\left(\pi\Pi\left(x'\right)\right)^{2}+\frac{1}{K}\left(\partial_{x'}\phi\right)^{2},\partial_{x}\phi\partial_{x}\phi_{0}\right]\\
&=i\frac{u}{2\pi}\int\mathrm{d}x'\left[K\left(\pi\Pi\left(x'\right)\right)^{2},\partial_{x}\phi\right]\partial_{x}\phi_{0}\\
&=i\frac{u}{2\pi}\int\mathrm{d}x'K\pi^{2}\left(-\partial_{x}\left(2i\delta\left(x-x'\right)\Pi\left(x'\right)\right)\right)\partial_{x}\phi_{0}\\
&=u\pi K\partial_{x}\Pi\left(x\right)\partial_{x}\phi_{0}
\end{align}
Together with its complementary term
\begin{align}
i\int\mathrm{d}x'\left[\partial_{x'}\phi\partial_{x'}\phi_{0},\mathcal{H}^{LL}\left(x\right)\right],
\end{align}
it yields a full derivative
\begin{align}
&i\int\mathrm{d}x'\Big\{\left[\mathcal{H}^{LL}\left(x'\right),\partial_{x}\phi\partial_{x}\phi_{0}\right]+\left[\partial_{x'}\phi\partial_{x'}\phi_{0},\mathcal{H}^{LL}\left(x\right)\right]\Big\} \nonumber \\
&=u\pi K\partial_{x}\left(\Pi\left(x\right)\partial_{x}\phi_{0}\right).
\end{align}
Once again, it is straightforward to deduce the corresponding contribution to $J_h$. The rest of the quadratic contributions to the heat current operator may be derived by a simple change of the field labels or by substitution
\begin{align}
&\phi\rightarrow\theta,\quad K\rightarrow\frac{1}{K}.
\end{align}

The last type of contributions we need to consider is the one coming from cosine terms like Eqs. ($\ref{eq:H_cos}$) and ($\ref{eq:H^ch_perp}$). Plugging into the commutator $i[H,\mathcal{H}]$, one encounters terms of the following  form
\begin{align}
&i\int\mathrm{d}x'\left[e^{i\theta\left(x'\right)},\left(\partial_{x}\phi\right)^{2}\right]=\\
&=\partial_{x}\phi i\int\mathrm{d}x'ie^{i\theta\left(x'\right)}i\pi\delta\left(x-x'\right) \\
&+i\int\mathrm{d}x'ie^{i\theta\left(x'\right)}i\pi\delta\left(x-x'\right)\partial_{x}\phi\\
&=-i\pi\left\{ \partial_{x}\phi,e^{i\theta\left(x\right)}\right\},
\end{align}
 which exactly cancels out with the complementary term $i\int\mathrm{d}x'\left[\left(\partial_{x'}\phi\right)^{2},e^{i\theta\left(x\right)}\right]$. Again, we can change the labels of the fields and substitute $\phi\rightarrow\theta$ to see that other combinations vanish too. The finite result for $J_h$ can be written in a matrix form as shown in Eq. (\ref{eq:HeatCurrent}).

\section{Evaluation of $\kappa_{xy}$ in the MSL and VBC phases}
\label{sec:ExpectationValues}

In this Appendix we present the derivation of key correlation functions, contributing to the calculation of heat current expectation value. We focus first on the MSL phase, where corrections to the quadratic bosonized form of the Hamiltonian Eq. (\ref{eq:H_original}) are irrelevant. The action corresponding to this quadratic part may be written in a block-matrix form:
\begin{align}
S&=\frac{T}{2 L}\sum_{\vec{q}}\Phi^{T}(-\vec{q})\begin{pmatrix}D_{1} & G_{1} & 0\\
G_{1}^{T} & \hat{D}_{0} & G_{-1}^{T}\\
0 & G_{-1} & D_{-1}
\end{pmatrix}\Phi(\vec{q})
\end{align}
where
\begin{align}
D_{1}&=\begin{pmatrix}\frac{q^{2}uK}{\pi} & i\frac{q\omega_{n}}{\pi}+gq^{2}\\
i\frac{q\omega_{n}}{\pi}+gq^{2} & \frac{q^{2}u}{\pi K}
\end{pmatrix}, \nonumber \\
G_{1}&=\begin{pmatrix}0 & g_{\perp,0}^{ch}\\
0 & g_{\perp}^{z}
\end{pmatrix}q^{2}\; ;
\end{align}
$D_{-1}$ and $G_{-1}$ can be obtained by taking $g\rightarrow-g$ and $g_{\perp,0}^{ch}\rightarrow -g_{\perp,0}^{ch}$, while $D_0$ by $u\rightarrow u_0$ and $K\rightarrow K_0$. Here $\Phi^{T}(\vec{q})$ are the Fourier components of the local field $\Phi^{T}$ defined after Eq. (\ref{eq:HeatCurrent}). In the chiral basis [Eq. (\ref{eq:BasisMSL})], the action acquires the form Eq. (\ref{eq:ActionMSL}) where the diagonal blocks are diagonalized. 
We then write the heat current operator [Eq. (\ref{eq:HeatCurrent})] in the same basis:
\begin{align}
\label{eq:HeatCurrentChiralBasis}
J_{h}\left(x\right)=\partial_{x}\Phi_{ch}^{T}\hat{J}_{h}^{ch}\partial_{x}\Phi_{ch} \; ,
\end{align}
where
\begin{align}
\hat{J}_{h}^{ch}=\begin{pmatrix}
\hat{Q}_{1} & \hat{K}_{1} & 0\\
\hat{K}_{1}^{T} & \hat{Q}_{0} & \hat{K}_{-1}^{T}\\
0 & \hat{K}_{-1} & \hat{Q}_{-1}
\end{pmatrix}
\end{align}
and
\begin{align}
&Q_{\pm1}=\frac{u^{2}}{4\pi K}
\begin{pmatrix}
-\left(1\mp\alpha\right)^{2} & 0\\
0 & \left(1\pm\alpha\right)^{2}
\end{pmatrix}, \nonumber  \\
&Q_{0}=\frac{u_{0}^{2}}{4\pi K_{0}}
\begin{pmatrix}
-1 & 0\\
0 & 1
\end{pmatrix}, \\
&\hat{K}_{1}=\frac{1}{8}
\begin{pmatrix}\{-ug_{\perp}^{z}-u_{0}g_{\perp}^{z}+g_{\perp,0}^{ch}u\} & \{ug_{\perp}^{z}-\frac{1}{8}u_{0}g_{\perp}^{z}-g_{\perp,0}^{ch}u\}\\
\{-ug_{\perp}^{z}+\frac{1}{8}u_{0}g_{\perp}^{z}-g_{\perp,0}^{ch}u \} & \{ ug_{\perp}^{z}+u_{0}g_{\perp}^{z}+g_{\perp,0}^{ch}u \}
\end{pmatrix}.\nonumber
\end{align}
Here $\alpha$ is related to $g$ via Eq. (\ref{alpha_g_def}), and $K_{-1}$ can be obtained by taking $g_{\perp,0}^{ch}\rightarrow -g_{\perp,0}^{ch}$.

To proceed with the calculation of the net heat current in the presence of a small temperature difference $\Delta T$ applied across the Kagome strip, we first consider the equilibrium contributions to $\langle J_h\rangle$ which include several correlation functions of the bosonic fields $\phi_l$. These are straightforwardly obtained by inverting the action Eq. (\ref{eq:ActionMSL}), and approximating the result up to second order in the inter-chain coupling constants $g_{\perp}^{z}$, $g_{\perp,0}^{ch}$, both assumed to be weak. This yields the following expressions:
\begin{widetext}
	\begin{align}
	\frac{T}{ L}\begin{pmatrix}\langle\phi_{\vec{q},0}^{R}\phi_{-\vec{q,}0}^{R}\rangle & \langle\phi_{\vec{q},0}^{R}\phi_{-\vec{q,}0}^{L}\rangle\\
	\langle\phi_{\vec{q},0}^{L}\phi_{-\vec{q,}0}^{R}\rangle & \langle\phi_{\vec{q},0}^{L}\phi_{-\vec{q,}0}^{L}\rangle
	\end{pmatrix}&\simeq\begin{pmatrix}G_{R,0} & 0\\
	0 & G_{L,0}
	\end{pmatrix}+A\begin{pmatrix}G_{R,0}^{2} & -G_{R,0}G_{L,0}\\
	-G_{R,0}G_{L,0} & G_{L,0}^{2}
	\end{pmatrix}\\
	\frac{T}{ L}\begin{pmatrix}\langle\phi_{\vec{q},1}^{R}\phi_{-\vec{q,}0}^{R}\rangle & \langle\phi_{\vec{q},1}^{R}\phi_{-\vec{q,}0}^{L}\rangle\\
	\langle\phi_{\vec{q},1}^{L}\phi_{-\vec{q},0}^{R}\rangle & \langle\phi_{\vec{q},1}^{L}\phi_{-\vec{q},0}^{L}\rangle
	\end{pmatrix}&\simeq\frac{q^{2}}{4}\times\begin{pmatrix}G_{R,1}G_{R,0}\left(-\frac{g_{\perp,0}^{ch}}{K}+g_{\perp}^{z}\right); & G_{R,1}G_{L,0}\left(\frac{g_{\perp,0}^{ch}}{K}-g_{\perp}^{z}\right)\\
	G_{L,1}G_{R,0}\left(-\frac{g_{\perp,0}^{ch}}{K}-g_{\perp}^{z}\right); & G_{L,1}G_{L,0}\left(\frac{g_{\perp,0}^{ch}}{K}+g_{\perp}^{z}\right)
	\end{pmatrix}\\
	\frac{T}{ L}\begin{pmatrix}\langle\phi_{\vec{q},1}^{R}\phi_{-\vec{q},1}^{R}\rangle & \langle\phi_{\vec{q},1}^{R}\phi_{-\vec{q},1}^{L}\rangle\\
	\langle\phi_{\vec{q},1}^{L}\phi_{-\vec{q},1}^{R}\rangle & \langle\phi_{\vec{q},1}^{L}\phi_{-\vec{q},1}^{L}\rangle
	\end{pmatrix}&\simeq\begin{pmatrix}G_{R,1} & 0\\
	0 & G_{L,1}
	 \end{pmatrix}+\left(\frac{q^{2}}{4}\right)^{2}\left(G_{R,0}+G_{L,0}\right)\times\\
	 &\begin{pmatrix}G_{R,1}^{2}\left(\frac{g_{\perp,0}^{ch}}{K}-g_{\perp}^{z}\right)^{2} & G_{R,1}G_{L,1}\left(\left(\frac{g_{\perp,0}^{ch}}{K}\right)^{2}-\left(g_{\perp}^{z}\right)^{2}\right)\\
	G_{R,1}G_{L,1}\left(\left(\frac{g_{\perp,0}^{ch}}{K}\right)^{2}-\left(g_{\perp}^{z}\right)^{2}\right) & G_{L,1}^{2}\left(\frac{g_{\perp,0}^{ch}}{K}+g_{\perp}^{z}\right)^{2}
	\end{pmatrix}
	\end{align}
	where
	\begin{align}
	 A&=\left(\frac{q^{2}}{4}\right)^{2}\left[\left(G_{R,1}+G_{L,-1}\right)
\left(\frac{g_{\perp,0}^{ch}}{K}-g_{\perp}^{z}\right)^{2}+\left(G_{L,1}+G_{R,-1}\right)\left(\frac{g_{\perp,0}^{ch}}{K}+g_{\perp}^{z}\right)^{2}\right]
	\end{align}
	and
	\begin{align}
	G_{R,1}^{-1}=\frac{q}{2K\pi}\left(qu_{-}-i\omega_{n}\right),\quad&G_{R,-1}^{-1}=\frac{q}{2K\pi}\left(qu_{+}-i\omega_{n}\right),\quad G_{R,0}^{-1}=\frac{q}{2K_{0}\pi}\left(qu_{0}-i\omega_{n}\right),\\
	G_{L,1}^{-1}=\frac{q}{2K\pi}\left(qu_{+}+i\omega_{n}\right),\quad&G_{L,-1}^{-1}=\frac{q}{2K\pi}\left(qu_{-}+i\omega_{n}\right),\quad G_{L,0}^{-1}=\frac{q}{2K_{0}\pi}\left(qu_{0}+i\omega_{n}\right).
	\end{align}
\end{widetext}
The rest of the correlation functions may be obtained by changing the chain index $1\rightarrow -1$  and substituting $g_{\perp,0}^{ch} \rightarrow -g_{\perp,0}^{ch}$. Direct correlations between the $l=1$ and $l=-1$ chains appear only to higher order in $g_{\perp}^{z}$, $g_{\perp,0}^{ch}$. Hence the leading contribution to $\langle J_h\rangle$ includes two types of terms, arising from the top side of the strip ($\langle\phi^\chi_1\phi^{\chi'}_1\rangle$, $\langle\phi^\chi_1\phi^{\chi'}_0\rangle$ where $\chi,\chi'=R,L$), and from the bottom part  ($\langle\phi^\chi_{-1}\phi^{\chi'}_{-1}\rangle$, $\langle\phi^\chi_{-1}\phi^{\chi'}_0\rangle$) separately. 
Accumulating these expressions, substituting in Eq. (\ref{eq:HeatCurrentChiralBasis}) and performing the summation over $\vec{q}=(\omega_n,q)$, we obtain a net heat current $\langle J_h\rangle$ as sum of two contributions which cancel at equilibrium. We then introduce a small temperature imbalance $\Delta T\ll T$ between the top and bottom sectors, assumed each to be at local equilibrium with temperature $T\pm \frac{1}{2}\Delta T$. We thus obtain a finite $\langle J_h\rangle=\kappa_{xy}\Delta T$, with $\kappa_{xy}$ an odd function of $g$ and $g_{\perp,0}^{ch}$ [Eq. (\ref{eq:HeatCurrentSL}) in the main text]. 

We next consider the VBC phase, where the bosonic fields are strongly interacting. However, an approximate free massive theory can still be employed, particularly in the vicinity of the special point $K=1/4$ where the Hamiltonian in each chain can be mapped to free fermions 
[Eq. (\ref{eq:FermionHamiltonian})]. The corresponding action is given by
\begin{align}
S_{f}\left[\psi_{R}^{\dagger},\psi_{R},\psi_{L}^{\dagger},\psi_{L}\right]=
\frac{T}{L}\sum_{\omega_{n},k}\Psi^{\dagger}\left(k,\omega_{n}\right)\hat{S}_{f}\Psi\left(k,\omega_{n}\right),
\end{align}
where
\begin{align}
&\Psi^{\dagger}\left(k,\omega_{n}\right)=\begin{pmatrix}\psi_{R}^{\dagger}\left(k,\omega_{n}\right) & \psi_{L}\left(-k,-\omega_{n}\right)\end{pmatrix}, \\
&\hat{S}_{f}=\begin{pmatrix}-i\omega_{n}+u_{_R}k & E\\
E & -i\omega_{n}-u_{_L}k
\end{pmatrix}.
\end{align}
Here $\omega_{n}$ is the fermionic Matsubara frequency, and we drop the chain index $\left(l\right)$ for the fields; in the outer chains ($l=\pm 1$) $u_{_R}=u(1\mp\alpha)$, $u_{_L}=u(1\pm\alpha)$ and in the middle chain ($l=0$) $u_{_R}=u_{_L}=u_0$. This leads directly to the correlation functions
\begin{align}
\label{eq:FermionCorFunc}
&\frac{T}{ L}\begin{pmatrix}\langle\psi_{R,\vec{k}}^{\dagger}\psi_{R,-\vec{k}}\rangle & \langle\psi_{R,\vec{k}}^{\dagger}\psi_{L,-\vec{k}}\rangle\\
\langle\psi_{L,\vec{k}}^{\dagger}\psi_{R,-\vec{k}}\rangle & \langle\psi_{L,\vec{k}}^{\dagger}\psi_{L,-\vec{k}}\rangle
\end{pmatrix}\\
&=\frac{1}{\det S_{f}}\begin{pmatrix}-i\omega_{n}-u_{_L}k & -E\\
-E & -i\omega_{n}+u_{_R}k
\end{pmatrix}, \nonumber
\end{align}
where 
\begin{align}
\det S_{f}=\left(-i\omega_{n}+\pi gk\right)^{2}-\left(uk\right)^{2}-E^{2}.
\end{align}

The calculation of $\kappa_{xy}$ proceeds following the same approach as in the MSL phase: since the chains are weakly coupled, we assume a local equilibrium in chain $l$ at temperature $T_l=T+\frac{l}{2}\Delta T$, and evaluate the corresponding contribution 
$\langle J_{f}^{\left(l\right)}\rangle$  to the net expectation value of $J_{h}$ [Eq. (\ref{eq:HeatCurrent_f})]. Note that here, unlike the MSL phase, a net heat current stemming from the difference in velocities $u_{_R}$, $u_{_L}$ is present in each of the outer chains $l=\pm 1$.  
For the upper chain ($l=1$), we obtain
\begin{align}
&\langle J_{f}^{\left(1\right)}\rangle=\frac{T_1}{ L}\sum_{\vec{q}}\Big\{ u^{2}\left(1-\alpha\right)^{2}q\langle\psi_{R,\vec{q}}^{\dagger}\psi_{R,\vec{q}}\rangle\\
&+u^{2}\left(1+\alpha\right)^{2}q\langle\psi_{L,\vec{q}}^{\dagger}\psi_{L,\vec{q}}\rangle\Big\}. \nonumber
\end{align}
Plugging in the correlation functions [Eq. (\ref{eq:FermionCorFunc})] and employing the low $T$ approximation $E\gg T$, we get the leading order expression 
\begin{align}
\langle J_{f}^{\left(1\right)}\rangle&\simeq e^{-\frac{E}{T_1}}\times\frac{2}{\pi}\sqrt{2\pi}\alpha\left(1+\alpha^{2}\right)\sqrt{E}T_1^{\frac{3}{2}}.
\end{align}
Combining the result with the lower ($l=-1$) chain contribution, with the substitution $T_1\rightarrow T_{-1}$ and $\alpha\rightarrow -\alpha$, leads to the final expression for the heat current [Eq. (\ref{eq:HeatCurrentDimerFinal})].


\begin{thebibliography}{235}
\expandafter\ifx\csname natexlab\endcsname\relax\def\natexlab#1{#1}\fi
\expandafter\ifx\csname bibnamefont\endcsname\relax
  \def\bibnamefont#1{#1}\fi
\expandafter\ifx\csname bibfnamefont\endcsname\relax
  \def\bibfnamefont#1{#1}\fi
\expandafter\ifx\csname citenamefont\endcsname\relax
  \def\citenamefont#1{#1}\fi
\expandafter\ifx\csname url\endcsname\relax
  \def\url#1{\texttt{#1}}\fi
\expandafter\ifx\csname urlprefix\endcsname\relax\def\urlprefix{URL }\fi
\providecommand{\bibinfo}[2]{#2}
\providecommand{\eprint}[2][]{\url{#2}}

\bibitem{Anderson1973} P.W. Anderson, Mater. Res. Bull {\bf 8}, 153 (1973).


\bibitem{balents-2010} L. Balents, Nature {\bf 464}, 199 (2010).

\bibitem{SavaryBalents2017} L. Savary and L. Balents, Rep. Prog. Phys {\bf 80}, 016502 (2017).

\bibitem{Starykh}
O. A. Starykh and L. Balents, Phys. Rev. Lett. {\bf 93}, 127202 (2004).


\bibitem{Iwase1996} H. Iwase, M. Isobe, Y. Ueda, and H. Yasuoka, Phys. Soc. Jpn. {\bf 65}, 2397 (1996).

\bibitem{Azuma1994} M. Azuma, Z. Hiroi, M. Takano, K. Ishida, and Y. Kitaoka, Phys. Rev. Lett. {\bf 73}, 3463 (1994).

\bibitem{Kageyama1999} H. Kageyama, K. Yoshimura, R. Stern, N. V. Mushnikov, K. Onizuka, M. Kato, K. Kosuge, C. P. Slichter, T. Goto, and Y. Ueda, Phys. Rev. Lett. {\bf 82}, 3168 (1999).

\bibitem{NT}
A. A. Nersesyan and A. M. Tsvelik, Phys. Rev. B {\bf 67} 024422 (2003).

\bibitem{Senthil2004}
T. Senthil, L. Balents, S. Sachdev, A. Vishwanath and M. P. A. Fisher, Phys. Rev. B {\bf 70} 144407 (2004).


\bibitem{DMRG}
S. Yan, D. A. Huse and S. R. White, Science {\bf 332}, 1173
(2011); S. Depenbrock, I. P. McCulloch and U. Schollw{\"o}ck, Phys. Rev.
Lett. {\bf 109}, 067201 (2012).

\bibitem{Assa2013} S. Capponi, V. R. Chandra, A. Auerbach and M. Weinstein, Phys. Rev. B {\bf 87}, 161118(R) (2013).




\bibitem{Hiroi2001} Z.Hiroi, M.Hanawa, N.Kobayashi, M.Nohara, H.Takagi, Y.Kato, and M.Takigawa, J. Phys. Soc. Jpn. {\bf 70}, 3377 (2001)


\bibitem{Ofer2006}
O. Ofer {\it et al}., arXiv:cond-mat/0610540.

\bibitem{Helton2007} J. S. Helton, K. Matan, M. P. Shores, E. A. Nytko, B. M. Bartlett, Y. Yoshida, Y. Takano, A. Suslov, Y. Qiu, J.-H. Chung, D. G. Nocera and Y. S. Lee, Phys. Rev. Lett. {\bf 98}, 107204 (2007).

\bibitem{Olariu}
A. Olariu, P. Mendels, F. Bert, F. Duc, J. C. Trombe, M. A. de Vries, and A. Harrison, Phys. Rev. Lett.
{\bf 100}, 087202 (2008).

\bibitem{Yamashita2008} S. Yamashita, T. Yamamoto, Y. Nakazawa, M. Tamura and R. Kato, Nature Phys. {\bf 4}, 459 (2008).

\bibitem{Okamoto2009} Y.Okamoto, H.Yoshida, and Z.Hiroi, J. Phys. Soc. Jpn. {\bf 78}, 033701, (2009).

\bibitem{Matan}
K. Matan et al., Phys. Rev. B {\bf 83}, 214406 (2011).


\bibitem{Yamashita2009} M. Yamashita, N. Nakata, Y. Kasahara, T. Sasaki, N. Yoneyama, N. Kobayashi, S. Fujimoto, T. Shibauchi and Y. Matsuda, Nature Phys. {\bf 5}, 44 (2009).

\bibitem{HirschbergerChinellLeeOng2015} M. Hirschberger, R. Chisnell, Y. S. Lee, and N. P. Ong, Phys. Rev. Lett. {\bf 115}, 106603 (2015).

\bibitem{HirschbergerKrizanCavaOng2015} M. Hirschberger, J. W. Krizan, R. J. Cava, N. P. Ong, Science {\bf 348}, 106 (2015).

\bibitem{WenWilczekZee1989} X. G. Wen, Frank Wilczek, and A. Zee, Phys. Rev. B {\bf 39}, 11413 (1989).

\bibitem{Baskaran1989} G. Baskaran, Phys. Rev. Lett. {\bf 63}, 2524 (1989).




\bibitem{KalmeyerLaughlin1987} V. Kalmeyer and R. B. Laughlin, Phys. Rev. Lett. {\bf 59}, 2095 (1987).

\bibitem{KalmeyerLaughlin1989} V. Kalmeyer and R. B. Laughlin, Phys. Rev. B {\bf 39}, 11879 (1989).




\bibitem{EranSela2015} G. Gorohovsky, R. G. Pereira, and E. Sela, Phys. Rev. B {\bf 91}, 245139 (2015).

\bibitem{Moessner2015} Y. C. He, S. Bhattacharjee, F. Pollmann, and R. Moessner, Phys. Rev. Lett. {\bf 115}, 267209 (2015).

\bibitem{TobiasMeng2015}  T. Meng, T. Neupert, M. Greiter, and R. Thomale, Phys. Rev. B {\bf 91} 241106(R) (2015).

\bibitem{Kitaev2006} A. Kitaev, Ann. Phys. {\bf 321}, 2 (2006).

\bibitem{KitaevSL}
G. Jackeli and G. Khaliullin, Phys. Rev. Lett. {\bf 102}, 017205 (2009).

\bibitem{Kasahara2018} Y. Kasahara, T. Ohnishi, Y. Mizukami, O. Tanaka, Sixiao Ma, K. Sugii, N. Kurita, H. Tanaka, J. Nasu, Y. Motome, T. Shibauchi and Y. Matsuda , Nature {\bf 559}, 227  (2018).

\bibitem{torque1}
K. A. Modic, B. J. Ramshaw, A. Shekhter and C. M. Varma, Phys. Rev. B {\bf 98}, 205110 (2018).

\bibitem{torque2}
S. D. Das, S. Kundu, Z. Zhu, E. Mun, R. D. McDonald, G. Li, L. Balicas, A. McCollam, G. Cao, J. G. Rau, H.-Y. Kee, V. Tripathi, S. E. Sebastian,
Phys. Rev. B {\bf 99}, 081101 (2019).

\bibitem{HyunyongJungPatrick} H. Lee, J. H. Han, and P. A. Lee, Phys. Rev. B {\bf 91}, 125413 (2015).

\bibitem{VinklerAvivRosch2018} Y. Vinkler-Aviv and A. Rosch, Phys. Rev. X {\bf 8}, 031032 (2018).

\bibitem{FQHladders}
M. Calvanese Strinati, E. Cornfeld, D. Rossini, S. Barbarino, M. Dalmonte, R. Fazio, E. Sela and L. Mazza, ``Laughlin-like states in bosonic and fermionic atomic synthetic ladders", Phys. Rev. X {\bf 7}, 021033 (2017).



\bibitem{Giamarchi} T. Giamarchi, \textit{Quantum Physics in One Dimension},
(Oxford, New York, 2004).

\bibitem{NLK93}
A. A. Nersesyan, A. Luther and F. V. Kusmartsev, Phys. Lett. A {\bf 176}, 363 (1993).

\bibitem{chirality}
Note that this choice of chiralities assumes $B/D>0$. Reversing either the sign of $B$ or $D$ simply interchanges $R$ and $L$.

\bibitem{GianarchiShulz1988} T. Giamarchi and H. J. Schulz, Phys. Rev. B, {\bf 37} 325 (1988)

\bibitem{OferMarciparChandraGazitPodolskyArovasKeren} O. Ofer, L. Marcipar, V. R. Chandra, S. Gazit, D. Podolsky, D. P. Arovas, and A. Keren, Phys. Rev. B {\bf 89}, 205116 (2014)


\end{thebibliography}
\end{document}